\documentclass[pre,twocolumn,aps,groupedaddress]{revtex4-2}
\usepackage[T1]{fontenc}
\usepackage[latin9]{inputenc}
\usepackage{array}
\usepackage{mathrsfs}
\usepackage{multirow}
\usepackage{amsmath}
\usepackage{amsthm}
\usepackage{amssymb}
\usepackage{mathbbol}
\usepackage{graphicx}
\usepackage{latexsym}
\usepackage{textcomp}
\usepackage{mathtools}
\usepackage{xcolor}
\usepackage[citecolor=magenta,colorlinks=true]{hyperref}
\usepackage{bm}
\definecolor{mycolor1}{rgb}{0.1, 0.6, 0.6}

\begin{document}

\title{Disorder Perturbation Expansion for Athermal Crystals}

\author{Pappu Acharya}
\email{pappuacharya@tifrh.res.in}
\affiliation{Centre for Interdisciplinary Sciences, Tata Institute of Fundamental Research, Hyderabad 500107, India}
\author{Debankur Das}
\email{debankurd@tifrh.res.in}
\affiliation{Centre for Interdisciplinary Sciences, Tata Institute of Fundamental Research, Hyderabad 500107, India}
\author{Kabir Ramola}
\email{kramola@tifrh.res.in}
\affiliation{Centre for Interdisciplinary Sciences, Tata Institute of Fundamental Research, Hyderabad 500107, India}

\date{\today}

\begin{abstract}
We introduce a perturbation expansion for athermal systems that allows an exact determination of displacement fields away from the crystalline state as a response to disorder. We show that the displacement fields in energy minimized configurations of particles interacting through central potentials with microscopic disorder, can be obtained as a series expansion in the strength of the disorder. We introduce a hierarchy of force balance equations that allows an order-by-order determination of the displacement fields, with the solutions at lower orders providing sources for the higher order solutions. This allows the simultaneous force balance equations to be solved, within a hierarchical perturbation expansion to arbitrary accuracy. We present exact results for an isotropic defect introduced into the crystalline ground state at linear order and second order in our expansion. We show that the displacement fields produced by the defect display interesting self-similar properties at every order. We derive a $|\delta r| \sim 1/r$ and $|\delta f| \sim 1/r^2$ decay for the displacement fields and excess forces at large distances $r$ away from the defect. Finally we derive non-linear corrections introduced by the interactions between defects at second order in our expansion. We verify our exact results with displacement fields obtained from energy minimized configurations of soft disks.
\end{abstract}

\maketitle

\section{Introduction}
\label{section_introduction}

Athermal materials represent a class of disordered systems where large scale properties are only weakly affected by ambient thermal fluctuations. Such behaviour emerges in many disordered systems when cooled to low temperatures~\cite{grigera1999observation}. Being governed purely by local constraints of mechanical equilibrium, athermal systems display amorphous disorder that arises from  the many possible arrangements of particles in minimum energy configurations. Such amorphous packings are inhomogeneous at the local scale and consequently are not described by the usual elasticity theories in continuum~\cite{landau1987theoretical}. Since the thermal motion of the constituent particles is irrelevant, athermal materials are not governed by fluctuation dissipation relations~\cite{kubo1966fluctuation}, and therefore offer interesting arenas to study non-equilibrium behaviour.
Many real-world systems can be classified as athermal, and they arise frequently in physics and biology. Examples include jammed packings of particles~\cite{jaeger1996granular,van2009jamming, henkes2007entropy,o2002random,o2003jamming,wyart2005rigidity, goodrich2012finite,ramola2017scaling,cates1998jamming,torquato2010jammed}, low temperature glasses ~\cite{berthier2011theoretical,kapteijns2019fast}, as well as densely packed tissues~\cite{bi2011jamming}. 

A fundamental type of disorder that arises in jammed athermal systems is polydispersity in particle sizes \cite{o2002random,Goodrich2014,tong2015crystals,acharya2020athermal}, which leads to amorphous energy minimized structures. This variation in particle sizes give rise to sources of localized stress within the material and are usually referred to as stress defects \cite{eshelby1956continuum,batra1987force}. Such defects play an important role in the bulk properties as well as the dynamics of amorphous systems. 
At the continuum scale stress defects have been modeled using bulk properties such as the elastic moduli of the medium and defect ~\cite{eshelby1957determination,duan2005eshelby},  however theories that incorporate microscopic disorder in the material are still missing. When the finite sizes of particles are considered, constructing coarse grained elasticity theories that account for disorder at the granular level becomes hard \cite{nampoothiri2020emergent}. 
At the local scale, the non-linearities involved in mechanical equilibrium conditions become important, as they have to be simultaneously satisfied by the entire configuration. This has made the characterization of athermal disorder and consequently the construction of statistical mechanical theories challenging~\cite{makse2002testing,tighe2010force,edwards1989theory,blumenfeld2009granular}.

Since the coupled non-linear equations of force balance at the microscopic scale are challenging to solve, many properties of athermal materials are only accessible numerically through sophisticated minimization procedures. 
Theoretical studies of force balanced networks have focused on diffusive models for scalar forces~\cite{liu1995force,coppersmith1996model}, as well as vectors along fixed lattice directions \cite{tighe2010force}. However incorporating force balance with continuous vectors within a microscopic model has remained elusive.
In this context, models of disordered athermal materials that can be exactly solved are important tools with which to understand athermal disorder as well as the sensitive deviations from linear elasticity that such materials display. A particularly appealing class of  systems are disordered athermal crystals where the periodicity of a reference configuration can be exploited ~\cite{acharya2020athermal,das2020long}, and consequently they offer a paradigm where exact results for disordered athermal systems can be obtained.

In this paper we develop a systematic perturbation expansion about the crystalline ordered state as a response to microscopic disorder. We introduce a hierarchy of simultaneous equations arising from microscopic force balance at each site, which can be solved at each order in the expansion. This allows an exact determination of the displacement fields that develop as a response to the disorder. We illustrate our techniques with an application to a model of polydispersed soft disks, where quenched disorder is introduced into the particle sizes, parametrized by a polydispersity parameter. We illustrate the convergence properties of such an expansion by a detailed analysis of a single defect introduced into a crystalline background. We also theoretically predict the displacement fields produced by a single defect in such a crystalline arrangement up to second order. We find that these displacement fields display a remarkable self-similar behaviour, with the solutions at higher orders in the perturbation theory resembling rescaled solutions at lower orders. To illustrate the non-trivial nature of the higher order terms in the perturbation expansion, we provide experimentally testable predictions arising from second order corrections, that are not captured by continuum elasticity theories. 


The plan of this paper is as follows. In Section \ref{section_greens_function_expansion} we develop a hierarchical perturbation expansion about the crystalline state, that allows us to exactly solve the equations of force balance and obtain the displacement fields to arbitrary accuracy.
In Section \ref{section_hierarchical} we examine the displacement fields generated by our expansion at linear order. We also examine the higher order solutions, providing exact equations up to second order. In Section \ref{section_deformable_disks} we contextualize our results in a model crystal composed of deformable disks, interacting through a paradigmatic soft-sphere Hamiltonian. Next, we examine the displacement fields produced by introducing a single defect into this system in Section \ref{section_single_defect}. We derive exact results for these displacement fields at second order.
We also examine the excess forces produced by this defect in Section \ref{section_forces_single_defect}.
In Section \ref{section_large_r} we derive the large lengthscale behaviour of the displacement and excess force fields produced by the single defect. We also discuss the universal features of our results, and their relation to continuum elasticity. Finally, in Section \ref{section_second_order_corrections}, we study the nature of the non-linear interactions between particles generated in our perturbation expansion, through the study of the displacement fields created by two defects in the crystalline system. We also provide an experimentally testable prediction which can be used to extract such non-linear interactions in athermal crystals. Finally, we conclude in Section \ref{section_conclusion} and provide avenues for further research.

\section{Athermal Crystals}
\label{section_athermal_crystals}

\begin{figure}[t!]
\centering
\includegraphics[width=0.85\linewidth]{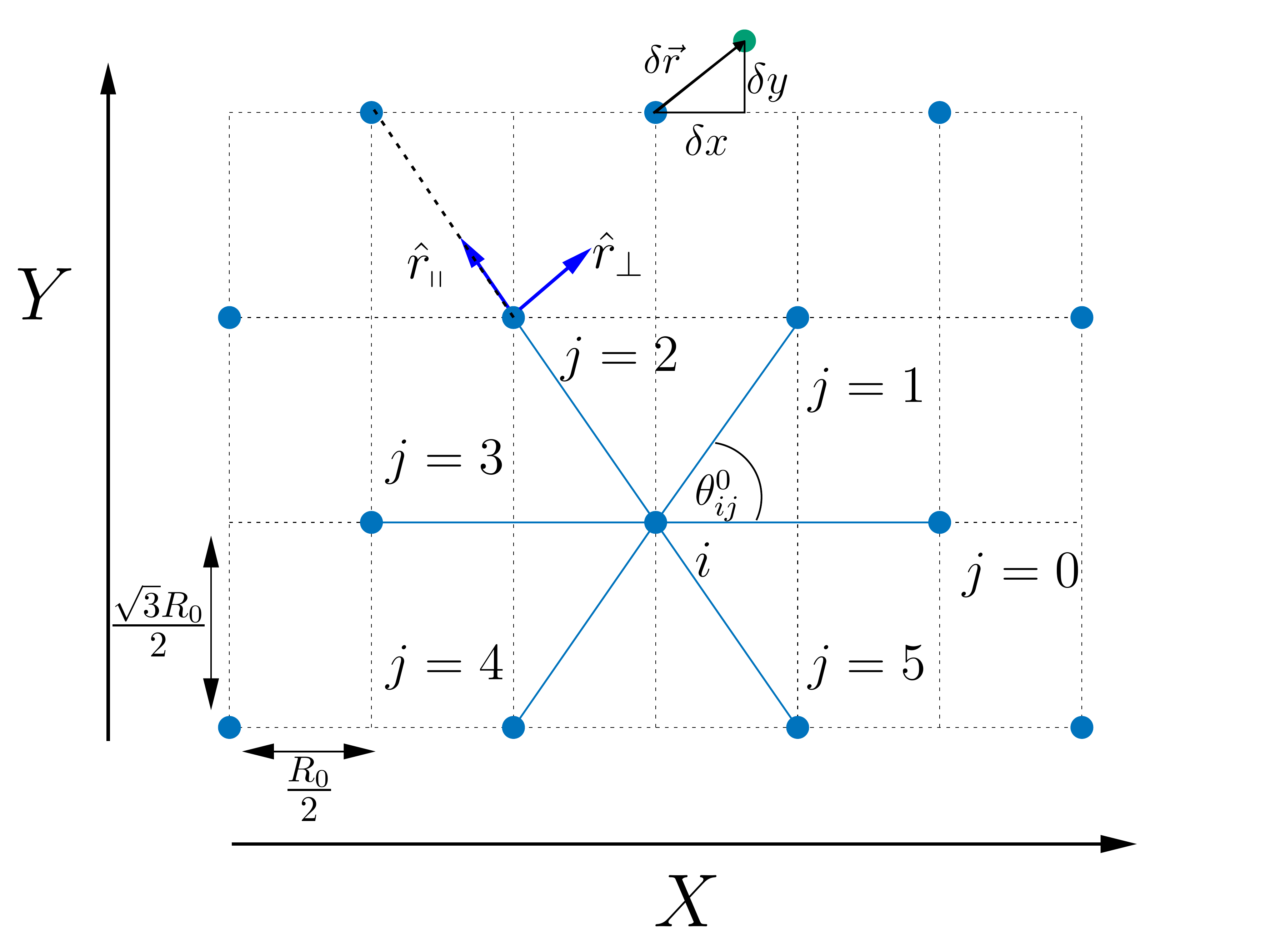}
\caption{The lattice convention used in this paper. We place the sites of the triangular lattice on a $2 L \times L$ rectangular grid, with the triangular lattice sites $i$ at position $\vec{r}_i \equiv (X,Y)$ such that $\text{mod}(X+Y,2)=0$. The lattice sites (particles) are represented by filled blue circles, and the lattice constant in the crystalline arrangement is $R_0$. Each particle $i$ has six neighbours which are labeled $j = 0$ to $5$. The bond angles between these particles $i$ and $j$ with respect to the $x$-axis are given by, $\theta_{ij}^{0} = \frac{2 \pi j}{6}$. As a response to the disorder, the particles are displaced to new locations (green circles). Here $\delta x$ and $\delta y$ represent the displacement of the particles from their original positions along the $x$ and $y$ directions respectively. The directions parallel $(\hat{r}_{\parallel})$ and orthogonal $(\hat{r}_{\perp})$ to the lattice directions are represented with blue arrows.
}
\label{fig_lattice_notation}
\end{figure}

The ground state of a system of particles interacting through short ranged central potentials is a triangular lattice arrangement \cite{theil2006proof}. The introduction of disorder can lead to fluctuations in the positions, and consequently elastic properties such as the local shear modulus~\cite{wang2018spatial,ma2016tailoring}.
Many studies have considered the effect of small disorder about the crystalline state, whose periodicity facilitates an exact theoretical treatment of their properties. However, when disorder is introduced into an athermal system, such as in amorphous configurations, developing theories that predict macroscopic properties becomes challenging ~\cite{zaccone2007theoretical,agnolin2008elastic,Goodrich2014}. 
In such crystals, the contribution of lattice phonons is suppressed, and crystallization results from energy minimization, and they exhibit properties of {\it athermal} materials.
Many examples of such disordered near-crystalline systems exist including glasses with high crystalline order \cite{watanabe2008direct,hu2017impact,tah2018glass}, polycrystals \cite{mecking1979deformation,sastry1971low} as well as colloidal crystals \cite{yiannourakou2010structural}.
In such systems, the changes in the interparticle potentials introduced through quenched disorder affects their large scale properties when force balance conditions are exactly imposed, leading to amorphous structures. This leads to elastic heterogeneity at the microscopic scale, and therefore continuum elasticity frameworks are inapplicable in such materials. 

In this work we consider {\it prestressed} crystals~\cite{acharya2020athermal}, where finite forces exist between the particles in the crystalline state. This is accomplished by the boundary conditions that confine the system. Examples include Wigner crystals, where particles interacting through central potentials crystallize under confinement at high densities~\cite{wigner1934interaction}, and jammed soft particle packings~\cite{Goodrich2014,tong2015crystals}.
We model such systems by considering a triangular lattice of particles with a lattice constant $R_0$ (see Fig.~\ref{fig_lattice_notation}). For ease of computation we place the sites of the triangular lattice on a $2 L \times L$ rectangular grid with the lattice sites $\vec{r}_i \equiv (X,Y)$ by maintaining $\text{mod}(X+Y,2)=0$~\cite{horiguchi1972lattice} as shown in Fig.~\ref{fig_lattice_notation}. Although, we use this convention in our computations, all the plotted figures represent our results on the actual triangular lattice. Thus the reciprocal lattice vector can be written as 
\begin{equation}
\vec{k} \equiv (k_x,k_y) \equiv \left( \frac{2 \pi l}{2 L}, \frac{2 \pi m}{L} \right),
\label{eq_reciprocal_definition}
\end{equation}    
and the volume of the system is $V = 2 L^2$. We also define fundamental lattice translation vectors $\vec{\mathbb{r}}_0 = (2,0)$, $\vec{\mathbb{r}}_1 = (1,1)$, $\vec{\mathbb{r}}_2 = (-1,1)$, $\vec{\mathbb{r}}_3 = (-2,0)$, $\vec{\mathbb{r}}_4 = (-1,-1)$, $\vec{\mathbb{r}}_5 = (1,-1)$.

\section{Disorder Perturbation Expansion}
\label{section_greens_function_expansion}

We begin by analyzing the response of the crystalline state to the presence of small amounts of disorder. The energy minimized configuration with no disorder in the microscopic potential is the triangular lattice arrangement of particles (see Fig.~\ref{fig_lattice_notation}). Our formulation considers any central force law with microscopic disorder, that causes a perturbation of the system away from the crystalline state.
We consider particles interacting through pairwise central potentials of the form
\begin{eqnarray}
u(\vec{r}_{ij},\{\zeta_{i}\}) = \mathcal{F}\left(| \vec{r}_{ij}|,\{\zeta_{i}\}\right).
\label{general_energy_law}
\end{eqnarray}
Here $\vec{r}_{ij} = \vec{r}_{j} - \vec{r}_{i}$ is the vector distance between the particles $i$ and $j$ located at positions $\vec{r}_i$ and $\vec{r}_j$ respectively. In this paper, we focus our discussion considering the pairwise central potential is short ranged. 
The variables $\{ \zeta_i \}$ represent quenched scalar variables at every site, which can be tuned to create pertubations about the crystalline state.
Additionally we assume that the bond energy $u(\vec{r}_{ij},\{\zeta_{i}\})$ associated with pairs of particles $i$ and $j$, is symmetric with respect to the scalar disorder at each site, i.e. it is a function of $\delta \zeta_{i} + \delta \zeta_{j}$.
In Sections \ref{section_deformable_disks} - \ref{section_large_r} we describe an application of our theory for such a potential, a system of polydispersed soft disks where the quenched variable $\{ \zeta_i \}$ is identified with the radii of the particles. 
The total energy is represented as a sum over all interacting pairs of particles, which can be expressed as a sum of their bond energies
\begin{equation}
U(\{\zeta_{i}\}) = \frac{1}{2}\sum_{ij} u(\vec{r}_{ij},\{\zeta_{i}\}).  
\end{equation}
Next, Eq.~(\ref{general_energy_law}) can be used to derive the interparticle forces as
\begin{equation}
\vec{f}_{ij}(\vec{r}_{ij}) = -\partial_{| \vec{r}_{ij}|} \mathcal{F} \left(| \vec{r}_{ij}|,\{\zeta_{i}\}\right) \hat{r}_{ij},
\label{general_force_law_equation}
\end{equation}
where $\hat{r}_{ij}$ is the unit vector along the interparticle distance vector. We can decompose the forces along the $x$ and $y$ directions as $\vec{f}=(f^{x},f^{y})$.
Crucially for central potentials the vector $\vec{f}_{ij}$ reverses sign under index interchange $i \leftrightarrow j$.

We focus specifically on states in mechanical equilibrium, i.e. energy minimized configurations.
Our formulation applies for attractive as well as repulsive potentials. Each configuration of the disordered system has contact forces $\{ \vec{f}_{ij} \}$ between particles $i$ and $j$.
The ground states of the system consist of configurations with each particle in force balance. We have
\begin{equation}
\sum_{j} f^{x}_{ij} = 0, ~~~\sum_{j} f^{y}_{ij} = 0, ~~~~\forall ~i.
\label{eq_force_balance_full}
\end{equation}
Here $f^{x(y)}_{ij}$ are the $x(y)$ components of the forces between particles $i$ and $j$, and the sum is over all particles $j$ in contact with particle $i$. 
In the next Section we use the perturbed solutions described above to solve these force balance equations at each order. The perturbation away from the crystal can be parametrized by the scalar variable at each site as
\begin{equation}
\zeta_i = \zeta_0 + \lambda\delta \zeta_i.
\end{equation}
Here, the variable $\lambda$ represents a tuning parameter that controls the strength of the quenched disorder in the system, $\lambda = 0$ corresponds to the crystalline state.

As a response to the presence of disorder, the positions of the particles deviate from their crystalline values $\vec{r}^{(0)}_i = \{x^{(0)}_{i}, y^{(0)}_{i}\}$ to a new mechanical equilibrium configuration $\vec{r}_i = \{x_{i}, y_{i}\}$. We define the vector {\it displacement field} $\{ \delta \vec{r}_i \}$ from the crystalline state at every site $i$ as
\begin{equation}
\vec{r}_{i} = \vec{r}^{(0)}_{i} + \delta \vec{r}_{i}.  
\end{equation}
A crucial aspect of our analysis is the decomposition of the displacement fields into their components along the fixed coordinate directions $x,y,...$ as $\{ \delta \vec{r}_i \} \equiv \{ \delta x_i, \delta y_i, ... \}$. This allows us to exactly determine the displacements in the presence of disorder. This can be accomplished since the perturbed positions can be used to generate the excess forces through the microscopic force law in Eq. (\ref{general_force_law_equation}). In $d$ dimensions, the $N d$ component displacement field vector can then be uniquely determined by the $N d$ equations of force balance in the system.
In this paper, we present results focused on two dimensional systems ($d = 2$), however our techniques can be extended to higher dimensional systems as well. 
The perturbed positions at each site $i$ can be expressed as
\begin{eqnarray}
\nonumber
x_{i} &=& x^{(0)}_{i} + \delta x_{i},\\
y_{i} &=& y^{(0)}_{i} + \delta y_{i}.
\label{eq_perturbed_positions}
\end{eqnarray}
The forces $\vec{f}_{ij} =(f^{x}_{ij},f^{y}_{ij})$ between any two particles $i$ and $j$ also change in response to this perturbation in the positions. We have
\begin{eqnarray}
\nonumber
f^{x}_{ij} &=& f^{x(0)}_{ij} + \delta f^{x}_{ij},\\
f^{y}_{ij} &=& f^{y(0)}_{ij} + \delta f^{y}_{ij},
\label{eq_perturbed_forces}
 \end{eqnarray}
where $\{ f^{x(0)}_{ij}, f^{y(0)}_{ij}\}$ are the forces along the $x$ and $y$ directions between particles $i$ and $j$ in the crystalline ordered state.
It is also convenient to define relative displacements between two neighbouring particles
\begin{eqnarray}
\delta x_{ij} &=& \delta x_{j} -\delta x_{i},\\
\nonumber
\delta y_{ij} &=& \delta y_{j} -\delta y_{i},
\label{eq_relative_displacements}
 \end{eqnarray}
as well as a corresponding sum involving the quenched random variables at neighbouring particles
\begin{eqnarray}
\delta \zeta_{ij} &=& \delta \zeta_{j} + \delta \zeta_{i}.
\label{eq_relative_sigma}
 \end{eqnarray}
Our analysis begins with the expansion of the interparticle forces in terms of the perturbed positions. The force $\vec{f}_{ij}$ between particles $i$ and $j$ can be Taylor expanded about its value in the crystalline ground state as 
\begin{eqnarray}
\nonumber
\delta f^x_{ij} &=& C_{ij}^{xx} \delta x_{ij} + C_{ij}^{xy} \delta y_{ij} + C_{ij}^{x \zeta} \delta \zeta_{ij} + \\
\nonumber
&& C_{ij}^{xxx} \delta x_{ij}\delta x_{ij} + C_{ij}^{xxy} \delta x_{ij}\delta y_{ij} + C_{ij}^{xyy} \delta y_{ij}\delta y_{ij}+\\
\nonumber
&& C_{ij}^{xx\zeta} \delta x_{ij}\delta\zeta_{ij} + C_{ij}^{xy\zeta}\delta y_{ij}
\delta\zeta_{ij}+ C_{ij}^{x\zeta\zeta} \delta\zeta_{ij}\delta\zeta_{ij}+ \hdots\\
\delta f^y_{ij} &=& C_{ij}^{yx} \delta x_{ij} + C_{ij}^{yy} \delta y_{ij} + C_{ij}^{y \zeta} \delta \zeta_{ij} + \\
\nonumber
&& C_{ij}^{yxx} \delta x_{ij} \delta x_{ij} + C_{ij}^{yxy} \delta x_{ij}\delta y_{ij} + C_{ij}^{yyy} \delta y_{ij}\delta y_{ij}+\\
\nonumber
&& C_{ij}^{yx\zeta} \delta x_{ij}\delta\zeta_{ij} + C_{ij}^{yy\zeta}\delta y_{ij}
\delta\zeta_{ij}+ C_{ij}^{y\zeta\zeta} \delta\zeta_{ij}\delta\zeta_{ij}+ \hdots
\end{eqnarray}

The coefficients $C_{ij}^{\alpha \beta}$ and $C_{ij}^{\alpha \beta\gamma}$ do not depend on the particle index $i$, as they are translationally invariant, being expressible purely in terms of the reference crystalline state. The general coefficient in the expansion at first order is given by
\begin{equation}
C_{ij}^{\alpha \beta} =- \frac{\partial \left( \mathcal{F} \left(\frac{r_{ij}^{\alpha}}{r_{ij}} \right) \right)}{\partial r^{\beta}}\Big|_{r_{ij}^{\beta(0)}},
\end{equation}
while the general coefficient at second order is given by
\begin{equation}
C_{ij}^{\alpha \beta \gamma} = -\frac{\partial^2 \left( \mathcal{F} \left( \frac{r_{ij}^{\alpha}}{r_{ij}} \right) \right)}{\partial r_{ij}^{\beta}\partial r_{ij}^{\gamma}}\Big|_{r_{ij}^{\beta(0)},r_{ij}^{\gamma(0)}}.
\end{equation}
Here $r_{ij} = \sqrt{\sum_{\alpha} r^{\alpha}_{ij} r^{\alpha}_{ij}}$ represents the magnitude of the distance vector $\vec{r}_{ij}$. The higher order coefficients can be obtained in a similar manner.
The interparticle forces, being drawn from the central potential have the form $f^{x(y)}_{ij} = \kappa(x_{ij},y_{ij},\zeta_{ij}) \frac{x(y)_{ij}}{r_{ij}}$ which is antisymmetric upon exchange of index $i \leftrightarrow j$. Therefore the terms at every order in the above force expansion should be antisymmetric with respect to this index interchange. Since $\delta x(y)_{ij}$ is antisymmetric under this interchange at all orders and $\delta \zeta_{ij} $ is symmetric, the $n^{th}$ order coefficients $C^{\alpha \beta \gamma ...}_{ij}$ is antisymmetric if the combined factors of $\delta x, \delta y$, and $\delta \zeta$ associated with the coefficient are symmetric, and vice versa.

Next, we expand the perturbed positions away from the crystalline state as an expansion in the disorder tuning parameter $\lambda$ as
\begin{eqnarray}
\nonumber
x_i &=& x^{(0)}_{i} + \lambda\delta x^{(1)}_{i} + \lambda^2\delta x^{(2)}_{i} + \lambda^3\delta x^{(3)}_{i}+ \hdots\\
y_i &=& y^{(0)}_{i} + \lambda\delta y^{(1)}_{i} + \lambda^2\delta y^{(2)}_{i} + \lambda^3\delta y^{(3)}_{i}+\hdots
\label{eq_displacement_expansions}
\end{eqnarray}
Here $\{\delta x^{(n)}_{i}, \delta y^{(n)}_{i}\}$ represent the $n^{th}$ order displacement fields, which are of magnitude $\mathcal{O}(\lambda^n)$.
We also define the relative displacements at every order
 \begin{eqnarray}
 \nonumber
 \delta x^{(1)}_{ij} &=& \delta x^{(1)}_{j} -\delta x^{(1)}_{i},\\
 \nonumber
 \delta y^{(1)}_{ij} &=& \delta y^{(1)}_{j} -\delta y^{(1)}_{i},\\
 \nonumber
 \delta x^{(2)}_{ij} &=& \delta x^{(2)}_{j} -\delta x^{(2)}_{i},\\
 \nonumber
 \delta y^{(2)}_{ij} &=& \delta y^{(2)}_{j} -\delta y^{(2)}_{i},\\
 &\vdots&
 \label{eq_relative_displacement_expansions}
 \end{eqnarray}
Similarly, the change in forces in Eq.~(\ref{general_force_law_equation}) can also be expressed as an expansion in the parameter $\lambda$, given by
\begin{eqnarray}
\nonumber
\delta f^x_{ij} &=& \lambda\delta f^{x(1)}_{ij} + \lambda^2\delta f^{x(2)}_{ij} + \lambda^3\delta f^{x(3)}_{ij} +\mathcal{O}(\lambda^4),\\
\delta f^y_{ij} &=& \lambda\delta f^{y(1)}_{ij} + \lambda^2\delta f^{y(2)}_{ij} + \lambda^3\delta f^{y(3)}_{ij} +\mathcal{O}(\lambda^4).
\label{eq_expansion_forces}
\end{eqnarray}
Inserting the expressions in Eq.~(\ref{eq_displacement_expansions}) and Eq.~(\ref{eq_relative_displacement_expansions}) into Eq.~(\ref{eq_expansion_forces}), and matching terms at first order, we obtain the change in interparticle forces at linear order along the $x$ and the $y$ directions. We have 
\begin{eqnarray}
\nonumber
\delta f^{x(1)}_{ij} &=& C_{ij}^{xx}\delta x^{(1)}_{ij} + C_{ij}^{xy} \delta y^{(1)}_{ij} + C_{ij}^{x \zeta} \delta \zeta_{ij},\\
\delta f^{y(1)}_{ij} &=& C_{ij}^{yx}\delta x^{(1)}_{ij} + C_{ij}^{yy} \delta y^{(1)}_{ij} + C_{ij}^{y \zeta} \delta \zeta_{ij}.
\label{eq_forcebal_first_order}
\end{eqnarray}
Similarly, matching terms at second order in Eq.~(\ref{eq_expansion_forces}), we arrive at the second order change in forces
\begin{small}
\begin{eqnarray}
\nonumber
\delta f^{x(2)}_{ij} &=& C_{ij}^{xx}\delta x^{(2)}_{ij}+ C_{ij}^{xy} \delta y^{(2)}_{ij} + C_{ij}^{xxx} \delta x^{(1)}_{ij}\delta x^{(1)}_{ij} + \\
\nonumber
&& C_{ij}^{xxy} \delta x^{(1)}_{ij}\delta y^{(1)}_{ij} +C_{ij}^{xyy} \delta y^{(1)}_{ij}\delta y^{(1)}_{ij} +C_{ij}^{xx\zeta} \delta x^{(1)}_{ij}\delta\zeta_{ij}+\\
\nonumber
&&C_{ij}^{xy\zeta}\delta y^{(1)}_{ij}
\delta\zeta_{ij} + C_{ij}^{x\zeta\zeta} \delta\zeta_{ij}\delta\zeta_{ij},\\
\nonumber
\delta f^{y(2)}_{ij} &=& C_{ij}^{yx}\delta x^{(2)}_{ij} + C_{ij}^{yy} \delta y^{(2)}_{ij} + C_{ij}^{yxx} \delta x^{(1)}_{ij}\delta x^{(1)}_{ij} +\\
\nonumber
&& C_{ij}^{yxy} \delta x^{(1)}_{ij}\delta y^{(1)}_{ij}+ C_{ij}^{yyy} \delta y^{(1)}_{ij}\delta y^{(1)}_{ij}  +C_{ij}^{yx\zeta} \delta x^{(1)}_{ij}\delta\zeta_{ij}+\\
&& C_{ij}^{yy\zeta}\delta y^{(1)}_{ij}
\delta\zeta_{ij} + C_{ij}^{y\zeta\zeta}
\delta\zeta_{ij}\delta\zeta_{ij},
\end{eqnarray}
\end{small}
The higher order terms in this perturbation expansion can be obtained in a systematic manner as described above, with the coefficients depending only on the higher order derivatives of the force law at $R_0$, interparticle separation in the reference crystalline state.

\section{Hierarchical Force Balance}
\label{section_hierarchical}

In this Section we use the perturbation expansion of the positions and forces developed above, to derive the displacements of the particle positions away from the crystalline ordered state. This is accomplished by imposing the simultaneous force balance conditions in Eq.~(\ref{eq_force_balance_full}) at every order. This provides enough equations to determine the displacement fields at each order. We show below that the displacement fields at every order can be determined uniquely within a hierarchical scheme, in which the solutions at lower orders act as ``sources'' that generate the displacement fields at higher orders.

\subsection{Linear Order}
\label{section_linear_order}
We begin by imposing the force balance conditions at first order in our expansion. Considering only terms up to linear order in Eq.~(\ref{eq_force_balance_full}), we arrive at the following set of equations representing force balance along the $x$ and $y$ directions respectively
\begin{eqnarray}
\nonumber
\sum_j \left(C_{ij}^{xx}\delta x^{(1)}_{ij} + C_{ij}^{xy}\delta y^{(1)}_{ij} + C_{ij}^{x\zeta}\delta \zeta_{ij}\right) &=& 0, ~\forall ~i, \\
\sum_j \left(C_{ij}^{yx}\delta x^{(1)}_{ij} + C_{ij}^{yy}\delta y^{(1)}_{ij} + C_{ij}^{y\zeta}\delta \zeta_{ij}\right) &=& 0, ~\forall ~i.
\label{1st_order_force}
\end{eqnarray}
where the coefficients $C_{ij}^{ \beta \gamma}$ depend only on the lattice constant $R_0$ and the neighbour $j$.
For a system of $N$ particles, there are $2N$ force balance constraints at every order. To calculate the displacement field for the entire system at a given order, one need to solve these $2N$ equations simultaneously, which requires the inversion of a $2N\times2N$ matrix. This is challenging for a system with a large number of particles. However, as the coefficients $C_{ij}^{\alpha \beta}$ are translationally invariant, a Fourier transform of the above equations helps to block diagonalize this matrix, into $2\times2$ blocks corresponding to each reciprocal lattice point $\vec{k}$. The Fourier transforms of the changes in positions and radii can be defined as
\begin{eqnarray}
\nonumber
&&\delta \tilde{x}^{(1)} (\vec{k}) = \sum_{\vec{r}} \exp(i \vec{k}. \vec{r}) \delta x^{(1)} (\vec{r}),\\
\nonumber
&&\delta \tilde{y}^{(1)} (\vec{k}) = \sum_{\vec{r}} \exp(i \vec{k}. \vec{r}) \delta y^{(1)} (\vec{r}),\\
&&\delta \tilde{\zeta} (\vec{k}) \hspace{0.05 in} =\hspace{0.05 in} \sum_{\vec{r}} \exp(i \vec{k}. \vec{r}) \delta \zeta (\vec{r}).
\end{eqnarray}
Here $\vec{r} \equiv i$ label the sites of the triangular lattice whereas $\vec{k}$ are the reciprocal lattice vectors defined in Eq.~(\ref{eq_reciprocal_definition}). It is also convenient to define the basic translation coefficients in Fourier space (using the notation in Fig.~\ref{fig_lattice_notation})
\begin{eqnarray}
\mathcal{F}_j(\vec{k}) &=& \exp(- i \vec{k}.\vec{\mathbb{r}}_j),
\end{eqnarray}
where $\vec{\mathbb{r}}_j$ are the fundamental lattice translation vectors.
Next, multiplying Eq.~(\ref{1st_order_force}) by $\exp(i \vec{k}. \vec{r})$ and summing over all sites $\vec{r} \equiv i$ leads to the following matrix equation at every reciprocal lattice point $\vec{k}$
\begin{equation}
\left(
\begin{matrix} 
A^{xx}(\vec{k}) & A^{xy}(\vec{k}) \\
A^{yx}(\vec{k}) & A^{yy}(\vec{k})
\end{matrix}
\right)
\left(
\begin{matrix} 
\delta \tilde{x}^{(1)} (\vec{k}) \\
\delta \tilde{y}^{(1)} (\vec{k})
\end{matrix}
\right)
= 
-\delta \tilde{\zeta }(\vec{k})
\left(
\begin{matrix} 
D^{x} (\vec{k}) \\
D^{y} (\vec{k})
\end{matrix}
\right).
\label{matrix eq_th}
\end{equation} 
We can interpret the matrix $A$ as an inverse Green's function of the system in Fourier space. The matrix elements of the inverse Green's function have the following explicit representations
 \begin{eqnarray}
\nonumber
A^{xx}(\vec{k}) &=& -\sum_{j=0}^{5} \left(1-\mathcal{F}_j(\vec{k})\right) C^{xx}_{ij},\\
\nonumber
A^{xy}(\vec{k}) &=&- \sum_{j=0}^{5} \left(1-\mathcal{F}_j(\vec{k})\right) C^{xy}_{ij},\\
\nonumber
A^{yx}(\vec{k}) &=& -\sum_{j=0}^{5} \left(1-\mathcal{F}_j(\vec{k})\right) C^{yx}_{ij},\\
A^{yy}(\vec{k}) &=&- \sum_{j=0}^{5} \left(1-\mathcal{F}_j(\vec{k})\right) C^{yy}_{ij}.
\end{eqnarray} 
Similarly we have 
\begin{eqnarray}
\nonumber
D^{x} (\vec{k}) &=& -\sum_{j=0}^{5} \left(1+\mathcal{F}_j(\vec{k})\right) C^{x\sigma}_{ij},\\
D^{y} (\vec{k}) &=& -\sum_{j=0}^{5} \left(1+\mathcal{F}_j(\vec{k})\right) C^{y\sigma}_{ij}.
\label{eq_D_functions}
\end{eqnarray}

\subsection{Response Green's Functions}

\begin{figure}[t!]
\hspace{-0.5cm}
\includegraphics[scale=0.25]{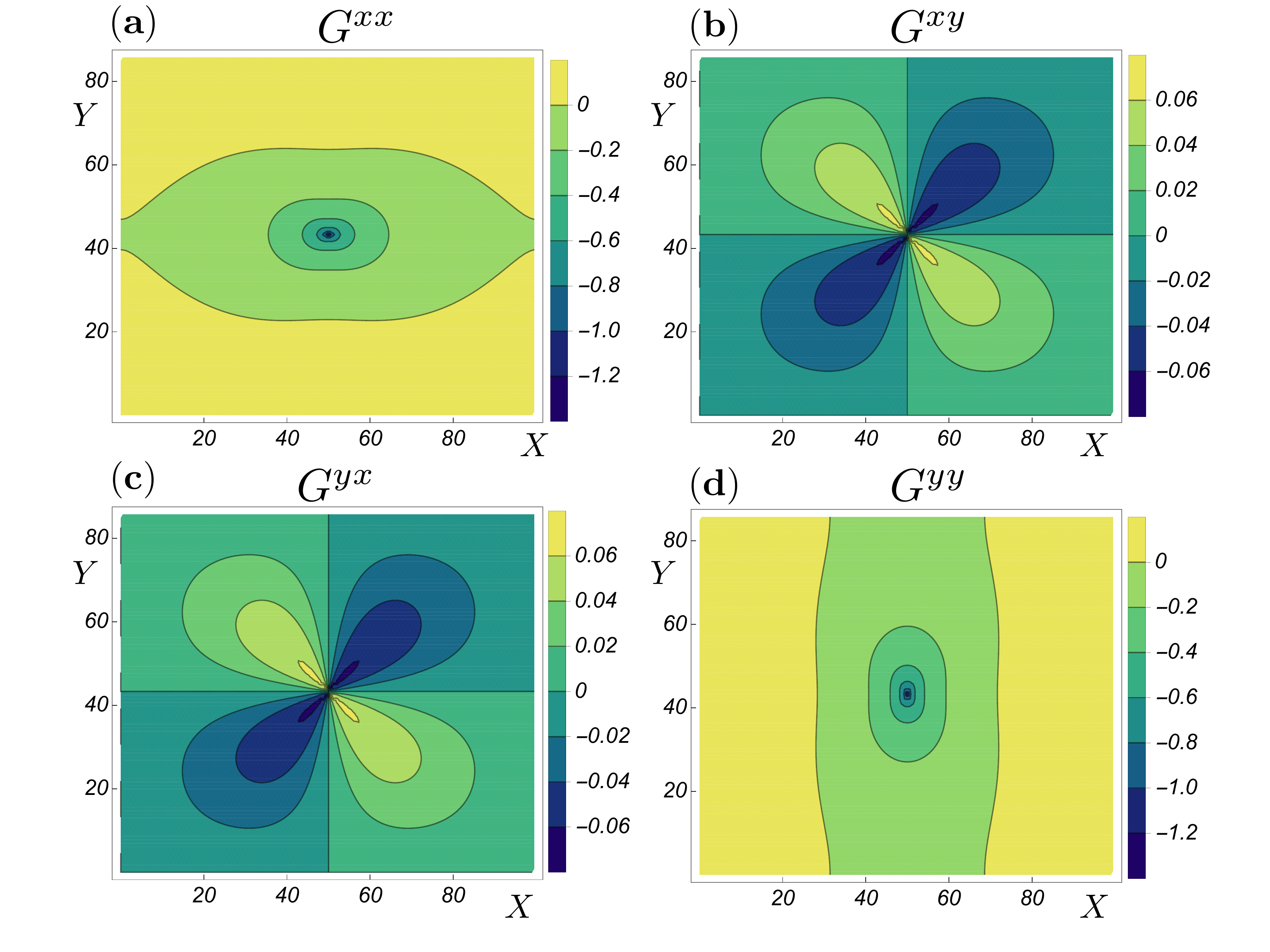}
\centering
\caption{The Green's functions of the response $G^{\mu \nu}(\vec{r})$ in real space. {\bf (a)} $G^{xx}(\vec{r})$ {\bf (b)} $G^{xy}(\vec{r})$, {\bf (c)} $G^{yx}(\vec{r})$, {\bf (d)} $G^{yy}(\vec{r})$. The transverse Green's functions ($G^{xy}$ and $G^{yx}$) have different limits in the $r\to \infty$ limit along different directions. All the Green's functions display a $\sim \log(|r|)$ dependence in the $r \to \infty$ limit. These Green's functions provide the solution to the displacement fields at all orders, with modified sources at each order. The behaviour of these functions generates the long distance behaviour of $\delta x(y)(\vec{r}) \sim \frac{1}{r}$ of the displacement fields away from a stress defect, for general pairwise potentials.}
\label{Fig_green}
\end{figure}

The Green's function can be expressed as a matrix
\begin{equation}
\tilde{G} = A^{-1} =\left(
\begin{matrix} 
\tilde{G}^{xx}(\vec{k}) & \tilde{G}^{xy}(\vec{k}) \\
\tilde{G}^{yx}(\vec{k}) & \tilde{G}^{yy}(\vec{k})
\end{matrix}
\right).
\label{green_mat}
\end{equation} 
The elements $\tilde{G}^{\mu \nu}$ can in turn be expressed in terms of the coefficients $A^{\mu \nu}$, which depend on the underlying potential through the coefficients $C_{ij}^{\alpha\beta}$. 
We next define source terms in Fourier space as
\begin{eqnarray}
\nonumber
\tilde{S}^{x(1)}(\vec{k}) &=& -D^{x}(\vec{k}) \delta \tilde{\zeta}(\vec{k}),\\
\tilde{S}^{y(1)}(\vec{k}) &=& -D^{y}(\vec{k}) \delta \tilde{\zeta}(\vec{k}).
\end{eqnarray}
Inverting Eq.~(\ref{matrix eq_th}), and using the above source, leads to an expression for the Fourier transformed displacement fields at first order in terms of the Fourier transform of the quenched disorder $\{\delta \tilde{\zeta}(\vec{k})\}$. We have
\begin{eqnarray}
\nonumber
\delta \tilde{x}^{(1)}(\vec{k})&=& \left(\tilde{G}^{xx}(\vec{k}) \tilde{S}^{x(1)}(\vec{k}) + \tilde{G}^{xy}(\vec{k}) \tilde{S}^{y(1)}(\vec{k}) \right),\\
\delta y^{(1)}(\vec{k}) &=& \left(\tilde{G}^{yx}(\vec{k}) \tilde{S}^{x(1)}(\vec{k}) + \tilde{G}^{yy}(\vec{k}) \tilde{S}^{y(1)}(\vec{k}) \right).
\label{dis_k_green}
\end{eqnarray}
The displacement fields at linear order can then be computed as an inverse Fourier transform, yielding
\begin{small}
 \begin{eqnarray}
\nonumber
 \delta x^{(1)}(\vec{r}) = \sum_{\vec{r}^\prime} G^{xx}(\vec{r}- \vec{r}^\prime)S^{x(1)}(\vec{r}^\prime) + \sum_{\vec{r}^\prime} G^{xy}(\vec{r}- \vec{r}^\prime)S^{y(1)}(\vec{r}^\prime),\\
 \nonumber
 \delta y^{(1)}(\vec{r}) = \sum_{\vec{r}^\prime} G^{yx}(\vec{r}- \vec{r}^\prime)S^{x(1)}(\vec{r}^\prime) + \sum_{\vec{r}^\prime} G^{yy}(\vec{r}- \vec{r}^\prime)S^{y(1)}(\vec{r}^\prime).\\
\label{eq_firstorder_disp_gen}
\end{eqnarray}
\end{small}
The real space Green's functions in the above expression are easily computed as an inverse transform of the Fourier space expressions in Eq.~(\ref{green_mat}) as
\begin{eqnarray}
\begin{aligned}
G^{\mu \nu}(\vec{r}) = \frac{1}{V}\sum_{\vec{k}}e^{-i \vec{k}.\vec{r}}\tilde{G}^{\mu \nu}(\vec{k}).
\end{aligned}
\end{eqnarray}
In Section \ref{section_large_r} we show that in the small $k$ limit, $\tilde{G}^{\mu \nu}(\vec{k}) \sim 1/|k|^2$, implying a $G^{\mu \nu}(\vec{r}) \sim \log(|r|)$ behaviour at large distances $\vec{r}$. These Green's functions in real space have been plotted in Fig.~\ref{Fig_green}. The functions $G^{xy}(\vec{r}) = G^{yx}(\vec{r})$ display a quadrupolar structure at large distances, as opposed to the sixfold symmetry of the underlying triangular lattice.

\subsection{Second Order}
\label{section_second_orders}

\label{sec_th}
Similarly, we can impose the force balance conditions at second order in our expansion, and extract the displacement fields as a solution of the corresponding simultaneous equations.
Considering only terms at second order in Eq.~(\ref{eq_force_balance_full}), we arrive at the second order force balance equations along the $x$ and $y$ directions respectively
 \begin{eqnarray}
 \nonumber
&&\sum_j\Bigl(C_{ij}^{xx}\delta x^{(2)}_{ij}+C_{ij}^{xy} \delta y^{(2)}_{ij} +C_{ij}^{xxx} \delta x^{(1)}_{ij}\delta x^{(1)}_{ij}\\
\nonumber
&&+C_{ij}^{xxy} \delta x^{(1)}_{ij}
\delta y^{(1)}_{ij} + C_{ij}^{xyy} \delta y^{(1)}_{ij}\delta y^{(1)}_{ij}
+ C_{ij}^{xx\zeta} \delta x^{(1)}_{ij}\delta\zeta_{ij}\\
\nonumber
&&+ C_{ij}^{xy\zeta}
 \delta y^{(1)}_{ij}\delta \zeta_{ij}
+ C_{ij}^{x\zeta\zeta} \delta\zeta_{ij}\delta\zeta_{ij} \Bigl)= 0,
~~~~\forall ~i,
\label{linearized_force_eq2}
\end{eqnarray}
\begin{eqnarray}
\nonumber
&&\sum_j\Bigl(C_{ij}^{yx}\delta x^{(2)}_{ij}+C_{ij}^{yy} \delta y^{(2)}_{ij} +C_{ij}^{yxx} \delta x^{(1)}_{ij}\delta x^{(1)}_{ij}\\
\nonumber
&&+C_{ij}^{yxy} \delta x^{(1)}_{ij}
\delta y^{(1)}_{ij} + C_{ij}^{yyy} \delta y^{(1)}_{ij}\delta y^{(1)}_{ij}
+ C_{ij}^{yx\zeta} \delta x^{(1)}_{ij}\delta\zeta_{ij}\\
\nonumber
&&+ C_{ij}^{yy\zeta}
 \delta y^{(1)}_{ij}\delta \zeta_{ij}
+ C_{ij}^{y\zeta\zeta} \delta\zeta_{ij}\delta\zeta_{ij} \Bigl)= 0,
~~~~\forall ~i,
\label{linearized_force_eq_sec}
\end{eqnarray}
Once again at second order we have $2N$ simultaneous equations to solve. From Eq.~(\ref{linearized_force_eq_sec}), it is clear that the first-order solutions serve as the source for the second order fields. These source terms in real space can be expressed as (with  $\vec{r} \equiv i$)
\begin{small}
\begin{eqnarray}
\nonumber
S_i^{x(2)}=\sum_j\Bigl(C_{ij}^{xxx} \delta x^{(1)}_{ij}\delta x^{(1)}_{ij} +C_{ij}^{xxy} \delta x^{(1)}_{ij}
\delta y^{(1)}_{ij} + C_{ij}^{xyy} \delta y^{(1)}_{ij}\delta y^{(1)}_{ij}\\
\nonumber
+ C_{ij}^{xx\zeta} \delta x^{(1)}_{ij}\delta\zeta_{ij}
+C_{ij}^{xy\zeta}
 \delta y^{(1)}_{ij}\delta \zeta_{ij}
+ C_{ij}^{x\zeta\zeta} \delta\zeta_{ij}\delta\zeta_{ij} \Bigl),\\
\nonumber
S_i^{y(2)}=\sum_j\Bigl(C_{ij}^{yxx} \delta x^{(1)}_{ij}\delta x^{(1)}_{ij} +C_{ij}^{yxy} \delta x^{(1)}_{ij}
\delta y^{(1)}_{ij} + C_{ij}^{yyy} \delta y^{(1)}_{ij}\delta y^{(1)}_{ij}\\
\nonumber
+ C_{ij}^{yx\zeta} \delta x^{(1)}_{ij}\delta\zeta_{ij}
+C_{ij}^{yy\zeta}
 \delta y^{(1)}_{ij}\delta \zeta_{ij}
+ C_{ij}^{y\zeta\zeta} \delta\zeta_{ij}\delta\zeta_{ij} \Bigl).\\
\label{eq_sec_source}
\end{eqnarray} 
\end{small}
The Fourier transforms of the second order source terms can be defined as
\begin{eqnarray}
\begin{aligned}
\tilde{S}^{x(2)}(\vec{k}) \hspace{0.05 in} =\hspace{0.05 in} \sum_{\vec{r}} \exp(i \vec{k}. \vec{r}) S^{x(2)}(\vec{r}).\hspace{0.16 in}\\
\tilde{S}^{y(2)}(\vec{k}) \hspace{0.05 in} =\hspace{0.05 in} \sum_{\vec{r}} \exp(i \vec{k}. \vec{r}) S^{y(2)}(\vec{r}).\hspace{0.16 in}
\label{eq_fourier_sec_source}
\end{aligned}
\end{eqnarray}
Using the above expressions, and following the same procedure as for linear order, we arrive at the following force balance equations in Fourier space at second order
\begin{equation}
\left(
\begin{matrix} 
A^{xx}(\vec{k}) & A^{xy}(\vec{k}) \\
A^{xy}(\vec{k}) & A^{yy}(\vec{k})
\end{matrix}
\right)
\left(
\begin{matrix} 
\delta \tilde{x}^{(2)} (\vec{k}) \\
\delta \tilde{y}^{(2)} (\vec{k})
\end{matrix}
\right)
= 
\left(
\begin{matrix} 
\tilde{S}^{x(2)} (\vec{k}) \\
\tilde{S}^{y(2)} (\vec{k})
\end{matrix}
\right).
\label{eq_sec_order_matrix}
\end{equation} 
We note that the matrix $A$ in the above equation is the same matrix that appears at first order in Eq.~(\ref{matrix eq_th}).
Next, we can invert  Eq.~(\ref{eq_sec_order_matrix}) and express the displacement fields using the Green's functions as
\begin{equation}
\left(
\begin{matrix} 
\delta \tilde{x}^{(2)} (\vec{k}) \\
\delta \tilde{y}^{(2)} (\vec{k})
\end{matrix}
\right)
= \left(
\begin{matrix} 
\tilde{G}^{xx}(\vec{k}) & \tilde{G}^{xy}(\vec{k}) \\
\tilde{G}^{xy}(\vec{k}) & \tilde{G}^{yy}(\vec{k})
\end{matrix}
\right)
\left(
\begin{matrix} 
\tilde{S}^{x(2)} (\vec{k}) \\
\tilde{S}^{y(2)} (\vec{k})
\end{matrix}
\right).
\end{equation} 
The second order displacement fields in Fourier space can then be written as
\begin{eqnarray}
\nonumber
\delta \tilde{x}^{(2)}(\vec{k})&=& \left(\tilde{G}^{xx}(\vec{k}) \tilde{S}^{x(2)}(\vec{k}) + \tilde{G}^{xy}(\vec{k}) \tilde{S}^{y(2)}(\vec{k}) \right),\\
\delta \tilde{y}^{(2)}(\vec{k}) &=& \left(\tilde{G}^{yx}(\vec{k}) \tilde{S}^{x(2)}(\vec{k}) + \tilde{G}^{yy}(\vec{k}) \tilde{S}^{x(2)}(\vec{k}) \right).
\label{dis_k_green_second}
\end{eqnarray}
We therefore arrive at the following expressions for the displacement fields at second order
\begin{small}
 \begin{eqnarray}
 \nonumber
 \delta x^{(2)}(\vec{r}) = \sum_{\vec{r}^\prime} \left[ G^{xx}(\vec{r}- \vec{r}^\prime)S^{x(2)}(\vec{r}^\prime) + G^{xy}(\vec{r}- \vec{r}^\prime)S^{y(2)}(\vec{r}^\prime) \right],\\
 \nonumber
 \delta y^{(2)}(\vec{r}) = \sum_{\vec{r}^\prime} \left[ G^{yx}(\vec{r}- \vec{r}^\prime)S^{x(2)}(\vec{r}^\prime) + G^{yy}(\vec{r}- \vec{r}^\prime)S^{y(2)}(\vec{r}^\prime) \right].\\
\label{eq_second_order_dis}
\end{eqnarray}
\end{small}
The Green's functions that appears above are the same as the Green's functions in the linear order solution in Eq.~(\ref{eq_firstorder_disp_gen}). However, unlike at first order, the second order solutions are expected to display non-linear effects arising from the disorder as the first order displacement fields  are coupled in the source terms in Eq.~(\ref{eq_sec_source}). We discuss such effects in detail in Section \ref{section_second_order_corrections}.

\subsection{Higher Orders}
\label{section_higher_orders}

The higher order terms in this expansion can now be analyzed in a similar manner. The same recursive scheme can be applied at higher orders, with the sources involving the solutions to the displacement fields at lower orders. For example, the equations of force balance at third order can be written in terms of the solutions at first and second order. Interestingly, due to the recursive structure of our perturbation expansion, the Green's function that appears at all orders are exactly the same. The source terms at higher orders couple the lower order solutions and hence display non-linear behaviour.



\section{Polydispersed Soft Disks}
\label{section_deformable_disks}


In this Section, we apply our perturbation expansion to a physically relevant system: soft disks with small polydispersity in particle sizes.
We begin with a triangular crystal composed of deformable disks interacting via one sided pairwise potentials that is now paradigmatic in the study of soft particles and deformable foams~\cite{durian1995foam,o2002random}. The form of the potential is
\begin{eqnarray}
\nonumber
V_{\sigma_{ij}}(\vec{r}_{ij}) &=& \frac{K}{\alpha}\left(1- \frac{| \vec{r}_{ij}|}{\sigma_{ij}}\right)^\alpha ~~\textmd{for}~~ r_{ij} < \sigma_{ij},\\
&=& 0 ~~~~~~~~~~~~~~~~~~~~~~\textmd{for}~~r_{ij} > \sigma_{ij}.
\label{energy_law}
\end{eqnarray}
Here $\vec{r}_{ij} = \vec{r}_{j} - \vec{r}_{i}$ is the vector distance between the particles $i$ and $j$ located at positions $\vec{r}_i$ and $\vec{r}_j$ respectively. $\sigma_{i j} = \sigma_i + \sigma_j$ is 
the sum of the radii $\sigma_i$ and $\sigma_j$ of the two particles. Here $K$ represents an overall stiffness constant, in the subsequent discussions we set $K = 1$ for simplicity. The interparticle forces are then given by
\begin{equation}
\vec{f}_{ij} = \frac{K}{\sigma_{ij}} \left(1- \frac{| \vec{r}_{ij}|}{\sigma_{ij}}\right)^{\alpha-1} \hat{r}_{ij},
\label{force_law_equation}
\end{equation}
where $\hat{r}_{ij}$ is the unit vector along the $\vec{r}_{ij}$ direction. In this work we present results for the harmonic potential, setting $\alpha = 2$. Our results can easily be generalized to other forms of the interparticle potential. When all the radii are equal, the minimum energy configuration of the system is the crystalline state with the particles forming a triangular lattice.  Choosing all the radii $\sigma_i$ to be equal to $\sigma_0$, the magnitude of the contact force in the crystalline state $f_0$ depends on the packing fraction $\phi = \sum_{i} \pi \sigma_i^2$ , and is given by $f_0 = \frac{1}{2\sigma_0}\left(1-\sqrt{\phi_c/\phi}\right)$. Here $\phi_c$ is the packing fraction of the marginal crystal with no overlaps between particles, and therefore no interparticle forces. This occurs in the hexagonally close packed structure, with $\phi_c = \pi/\sqrt{12}\approx 0.9069$. 
It is also convenient to define a parameter $\epsilon$, that quantifies the overcompression of the system
\begin{equation}
\epsilon = 1-R_0 = 1- \sqrt{\frac{\phi_c}{\phi}}.
\label{eq_epsilon_definition}
\end{equation}
The quantity $\epsilon$ therefore defines the lengthscale of overlap between particles in the pure crystalline arrangement.
Next, a perturbation away from the crystalline state can be caused by a change in the radii at each site
\begin{equation}
\sigma_i = \sigma_0 + \lambda\delta \sigma_i.
\end{equation}
Therefore, we can identify the radii of the particles as the quenched random variables that appear in the perturbation expansion in Section \ref{section_greens_function_expansion}. We have
\begin{equation}
\zeta \equiv \sigma.
\end{equation}

\begin{figure*}[t!]
\includegraphics[scale=0.35]{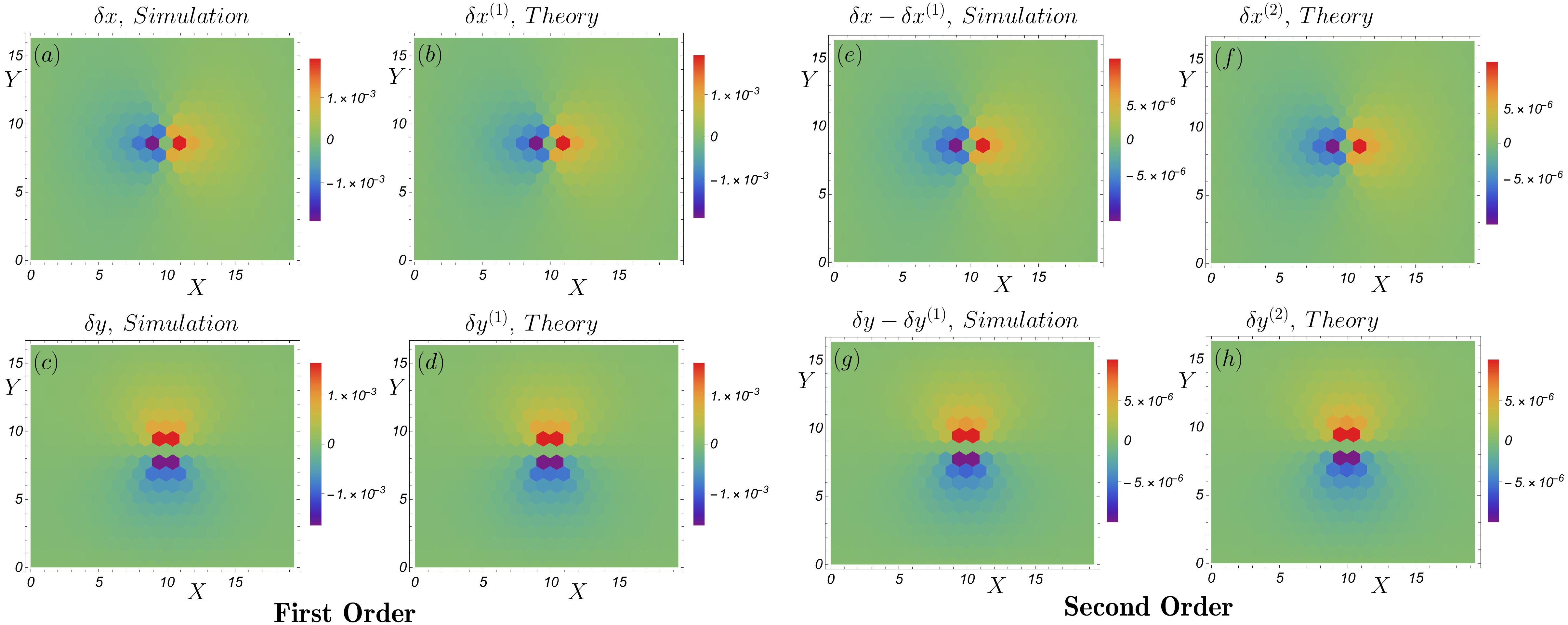}
\centering
\caption{Displacements of each grain away from their crystalline positions, produced by a single defect. Here each point in the $xy$-plane represents the original crystalline position of the grain. {\bf (a)} The displacement of each grain along the $x$ direction obtained from the simulations, {\bf (b)} The displacement of each grain along the $x$ direction obtained from the perturbation expansion at linear order. {\bf (c)} The displacement of each grain along the $y$ direction obtained from the simulations, {\bf (d)} The displacement of each grain along the $y$ direction obtained from the perturbation expansion at linear order. {\bf (e)} Displacement fields along the $x$ direction $\delta x - \delta x^{(1)}$ with the linear order solutions subtracted from the displacements obtained from simulations, {\bf (f)} the displacement of each grain along the $x$ direction obtained from the perturbation expansion at second order. {\bf (g)} Displacement fields along the $y$ direction $\delta y - \delta y^{(1)}$ with the linear order solutions subtracted from the displacements obtained from simulations, {\bf (h)} the displacement of each grain along the $y$ direction obtained from the perturbation expansion at second order. Note the difference in the magnitudes between the solutions at first and second order.}
\label{Fig_displacement_field} 
\end{figure*} 
The displacement fields in this system can then be derived using the techniques presented in Section \ref{section_greens_function_expansion}, 
For the harmonic case ($\alpha = 2$) the coefficients $C_{ij}^{\alpha \beta}$ that appear in the Green's function have particularly simple forms, and are given in the Appendix.
The Green's functions for this system in Fourier space are given by
\begin{eqnarray}
\begin{aligned}
\tilde{G}^{xx}(\vec{k}) = \Gamma_1 \frac{ (1-\epsilon)}{\Gamma_1\Gamma_2 -\Gamma^2_3} ,\\
\tilde{G}^{xy}(\vec{k}) = \Gamma_3 \frac{ (1-\epsilon)}{\Gamma_1\Gamma_2 -\Gamma^2_3}, \\
\tilde{G}^{yx}(\vec{k}) = \Gamma_3 \frac{ (1-\epsilon)}{\Gamma_1 \Gamma_2 -\Gamma^2_3}, \\
\tilde{G}^{yy}(\vec{k}) = \Gamma_2 \frac{ (1-\epsilon)}{\Gamma_1 \Gamma_2 -\Gamma^2_3}, 
\label{eq_exact_green}
\end{aligned}
\end{eqnarray}
where the functions $\Gamma_1$, $\Gamma_2$, $\Gamma_3$ have the following forms
\begin{eqnarray}
\nonumber
\Gamma_1 &=& (3-4 \epsilon ) \cos (k_x) \cos (k_y)-2 \epsilon  \cos (2 k_x)+6 \epsilon -3,\\
\nonumber
\Gamma_2 &=& (1-4 \epsilon ) \cos (k_x) \cos (k_y)+2 ( 1-\epsilon) \cos (2 k_x)+6 \epsilon -3,\\
\Gamma_3 &=&\sqrt{3}\sin(k_x)\sin(k_y).
\end{eqnarray}

\section{Single defect: series solution for displacement fields}
\label{section_single_defect}

In this Section, we use the perturbation theory developed in Sections \ref{section_greens_function_expansion} and \ref{section_hierarchical} to compute the displacement fields generated by the presence of a single defect in the crystalline background. To model such a situation, we increase the radius of a single particle by an amount $\delta\sigma$ at a position $\vec{R}$ in the crystalline arrangement. This causes an excess outward displacement $\delta \vec{r} \equiv \{ \delta x,\delta y \}$ of each grain as a response to the increase in the radius of the defect particle. Such inhomogeneities in materials are termed stress defects, since the defect causes an outward stress away from it. In the continuum elasticity framework, the stress and displacement are related through a stress geometry equation involving the elastic moduli, which can then be used to solve for a displacement field. In our case of a lattice of finite sized particles, the displacements give rise to forces through a microscopic force law, which through force balance conditions leads to a unique solution for the displacement fields. In addition, since we solve the equations of force balance simultaneously, our formulation can also be used to incorporate disorder at the microscopic scale.

In continuum elasticity, the displacement fields $\vec{u} \equiv \delta \vec{r}$ arising due to localised excess forces can be obtained using the well-known Lam\'e equation, which in two dimensions is~\cite{landau1987theoretical} 
\begin{eqnarray}
\mu \nabla^2 \vec{u}(\vec{r}) + (\lambda + \mu) \vec{\nabla}(\vec{\nabla} \cdot \vec{u}(\vec{r})) = -\vec{f}(\vec{r}),
\label{eq_lame}
\end{eqnarray}
where $\mu$ and $\lambda$ are the Lam\'e coefficients derived from the elastic moduli of the solid. The field $\vec{f}(\vec{r})$ represents the body force acting on the system at position $\vec{r}$.
Using the known elastic moduli for the triangular geometry, one can therefore obtain the displacement fields in the presence of external forces using Eq.~(\ref{eq_lame}). The generalization of this procedure to the case of a single stress defect can be obtained by solving the homogeneous case of Eq.~(\ref{eq_lame}) in a cylindrical geometry with a uniform normal force applied at the inner boundary. However, it as yet unclear how such Lam\'e equations emerge from a {\it microscopic} treatment of the finite-size of the particles, which can cause sensitive corrections to the continuum predictions. 
Our exact framework therefore provides an interesting route to verify the isotropic elasticity of a medium composed of finite-sized units. 
We show that the case of a triangular lattice arrangement yields displacement fields which display the underlying lattice symmetries at short lengthscales. However, in Section \ref{section_large_r} we show that the exact displacement fields at large distances away from a single defect obey the above continuum equations. The linear displacement fields along with Eq.~(\ref{eq_lame}) enables us to derive the macroscopic elastic moduli in terms of the coefficients obtained from the underlying interpaticle potential. The higher order solutions in our expansion can therefore be used to predict non-linear corrections to continuum elasticity that arise in athermal systems.

\subsection{Linear Order}

We begin by placing a single defect particle with a larger radius $\sigma_0 +\delta\sigma$ at the origin. The quenched disorder field is therefore non-zero only at the single site, and is given by
\begin{equation}
\sigma(\vec{r}) = \sigma_0 +\delta\sigma\delta(\vec{r}),  
\label{eq_site_disorder}
\end{equation}
leading to a constant source field in Fourier space
$\delta \Tilde{\sigma} (\vec{k}) = \delta \sigma$.
We first derive the displacement fields generated by the defect at the origin using the perturbation theory at linear order. It has been shown recently that mechanical equilibrium at linear order leads to non-trivial effects on fluctuations and correlations in athermal near-crystalline systems \cite{acharya2020athermal,das2020long}.
The field in Eq.~(\ref{eq_site_disorder}) can be used as a source in Eq.~(\ref{eq_firstorder_disp_gen}).
This yields the displacement fields at linear order $\{ \delta x^{(1)}, \delta y^{(1)} \}$. We next compare these with displacement fields obtained from a numerically minimized configuration of a single defect in the crystalline background. In Figs.~\ref{Fig_displacement_field}~{\bf (a)} and \ref{Fig_displacement_field}~{\bf (c)} we plot the $x$-displacement fields and $y$-displacement fields obtained from the numerical simulations. We also plot the linear order displacement fields $\delta x^{(1)}$ and $\delta y^{(1)}$ obtained from the perturbation expansion using Eq.~(\ref{eq_firstorder_disp_gen}) in Figs.~\ref{Fig_displacement_field}~{\bf (b)}, and \ref{Fig_displacement_field}~{\bf (d)}. The displacement fields obtained from our numerical minimization match exactly with the displacement fields computations at linear order. The difference between these two arises in the second order terms in our perturbation expansion.

\subsection{Second Order}
We next study the {\it difference} between the displacements obtained from our numerical minimization using the FIRE algorithm, and the predictions of our perturbation theory at linear order.
We compute the second order displacement fields along the $x$ direction $\delta x^{(2)}$ and $y$ direction $\delta y^{(2)}$ using Eq.~(\ref{eq_second_order_dis}). This involves the displacement fields at first order $\{ \delta x^{(1)}, \delta y^{(1)} \}$, which can be used to create the source term at second order using Eq.~(\ref{eq_sec_source}). The coefficients $C_{ij}^{\alpha \beta \gamma}$ that appear in these source terms are given in the Appendix.
This yields the displacement fields at second order $\{ \delta x^{(2)}, \delta y^{(2)} \}$, using the Green's function formalism in Eq.~(\ref{eq_second_order_dis}). 
We next compare these theoretical results at second order, with the difference between the displacement fields obtained from numerical minimization and the theoretically computed linear order displacements $\{ \delta x - \delta x^{(1)}, \delta y - \delta y^{(1)}\}$.
These are plotted in Figs.~\ref{Fig_displacement_field}~{\bf (e)} and \ref{Fig_displacement_field}~{\bf (g)}. The theoretical predictions of the second order displacement field is displayed in Figs.~\ref{Fig_displacement_field}~{\bf (f)} and \ref{Fig_displacement_field}~{\bf (h)}. The displacement fields obtained from numerical minimization match with our theoretical predictions at second order exactly. The difference between these two arises in the third order terms in our perturbation expansion. Remarkably, these second order displacement fields are very similar to the displacement fields obtained at first order, with a scaling factor $\mathcal{O}(\delta \sigma)$. We discuss the origin of such behaviour in our perturbation expansion in Section \ref{section_large_r}.

\subsection{Third Order}
\begin{figure}[t!]
\includegraphics[scale=0.38]{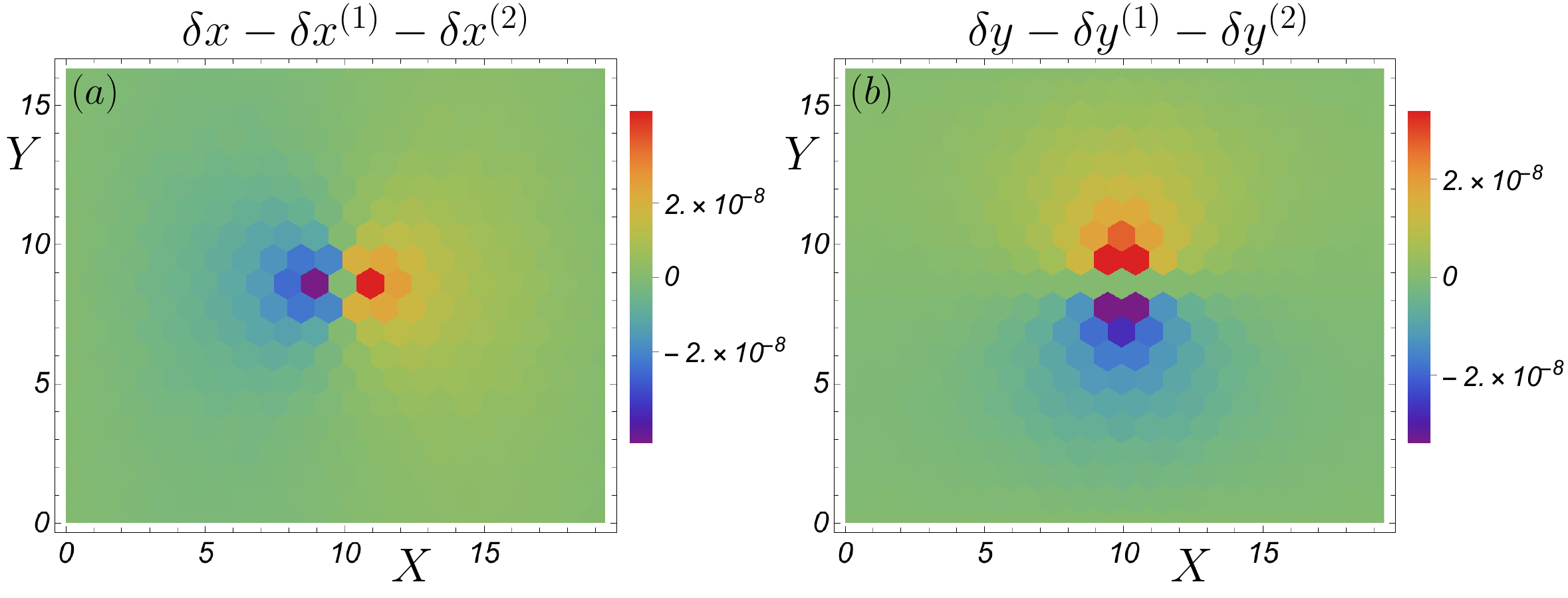}
\centering
\caption{Displacement fields produced by a single defect at third order, obtained by subtracting the first and second order solutions of the perturbation expansion from the numerically obtained displacements {\bf (a)} along the $x$ direction $\delta x - \delta x^{(1)} - \delta x^{(2)}$ and {\bf (b)} along the $y$ direction $\delta y - \delta y^{(1)} - \delta y^{(2)}$. These displacement fields resemble the first and second order solutions, with a rescaling factor.}
\label{displacement_field_third} 
\end{figure}

Following our method in Section \ref{section_second_orders}, the equations of force balance at third-order can next be written in terms of the solutions at lower orders. Once again, we can obtain these solutions, using the Green's function formalism, with the sources at third order involving the displacement fields at first and second order. This procedure is straightforward, and we have not attempted to extract the displacement fields at third order theoretically.
Instead, we have numerically extracted the displacement fields at third order by subtracting the theoretically obtained lower order solutions $\delta x(y)^{(1)}$ and $\delta x(y)^{(2)}$ from the actual displacements $\delta x(y)$ obtained in the numerically minimized configuration with a single defect. The quantities $\delta x - \delta x^{(1)} - \delta x^{(2)}$  and $\delta y - \delta y^{(1)} - \delta y^{(2)}$, which represent the third order solutions, are plotted in  Fig.~\ref{displacement_field_third} {\bf(a)} and Fig.~\ref{displacement_field_third} {\bf(b)} respectively.
Once again, we find that the third order displacement fields produced by a single defect in our perturbation expansion display very similar behaviour to the first order displacement fields, with a scaling factor $\mathcal{O}(\delta \sigma^2)$.

\section{Forces produced by a defect}
\label{section_forces_single_defect}

We next analyze the change in interparticle forces caused by the presence of a single defect in the system. Since the force at each bond depends on the displacements of the corresponding particles, we can use the displacement fields as predicted from our theory to calculate the interparticle forces at all orders. 
The change in interparticle forces between particles $i$ and $j$ can be computed at linear order from the expressions in Eq.~(\ref{1st_order_force}). We have
\begin{eqnarray}
\label{force_deviation}
\nonumber
\delta f^{x(1)}_{ij} &=& \frac{1}{V}\sum_{\vec{k}} \bigg(\Big(C^{xx}_{ij} \delta \tilde{x}^{(1)}(\vec{k}) +C^{xy}_{ij} \delta \tilde{y}^{(1)}(\vec{k})
\Big) \big(1-\mathcal{F}_{j}(\vec{k})\big)  \\
\nonumber
&&+ C^{x\sigma}_{ij}\big(1+\mathcal{F}_{j}(\vec{k}) \big) \delta\tilde{\sigma}(\vec{k}) \bigg)  \exp (-i\vec{k}.\vec{r}_i),\\
\nonumber
\delta f^{y(1)}_{ij} &=& \frac{1}{V}\sum_{\vec{k}} \bigg(\Big(C^{yx}_{ij} \delta \tilde{x}^{(1)}(\vec{k}) +C^{yy}_{ij} \delta \tilde{y}^{(1)}(\vec{k})
\Big) \big(1-\mathcal{F}_{j}(\vec{k})\big)  \\
&&+ C^{y\sigma}_{ij}\big(1+\mathcal{F}_{j}(\vec{k}) \big) \delta\tilde{\sigma}(\vec{k}) \bigg)  \exp (-i\vec{k}.\vec{r}_i).
\label{greenforce}
\end{eqnarray}

\begin{figure}[t!]
\hspace*{-0.5cm}
\includegraphics[scale=0.33]{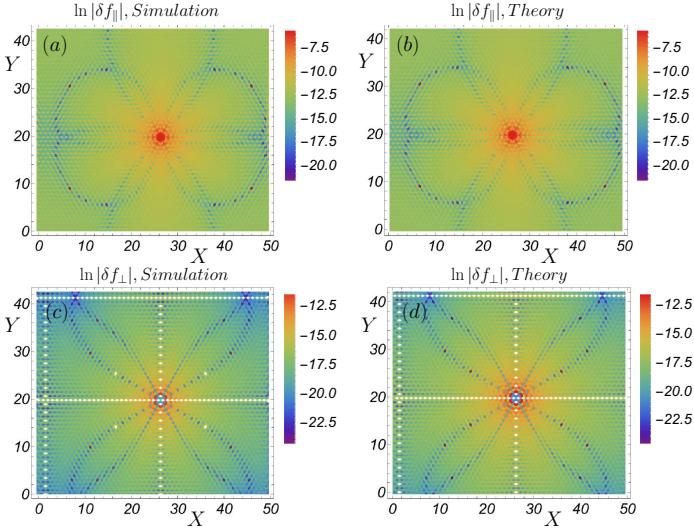}
\caption{The components of the interparticle forces parallel to the original lattice directions $f_{\parallel}$ and perpendicular to the lattice directions $f_{\perp}$. {\bf (a)} Plot of $\ln(|\delta f_\parallel|)$ at each bond produced by a single defect plotted at their original crystalline positions obtained from simulations. {\bf (b)} $\ln(|\delta f_\parallel|)$ at each bond produced by a single defect obtained from the perturbation expansion at linear order. {\bf (c)} $\ln(|\delta f_\perp|)$ at each bond produced by a single defect obtained from simulations. {\bf (d)} $\ln(|\delta f_\perp|)$ at each bond produced by a single defect obtained from the perturbation expansion at linear order. Note the order of magnitude difference in $\delta f_{\perp}$ and $\delta f_{\parallel}$. 
}
\label{Single_force_field}
\end{figure}

In order to better elucidate the nature of the forces in this athermal crystalline system, we define the parallel and perpendicular components of the interparticle forces with respect to the {\it original} lattice directions $\hat{r}_{\parallel}$ and $\hat{r}_{\perp}$ as $f_{\parallel}$ and $f_{\perp}$ (see Fig.~\ref{fig_lattice_notation}).
The deviation of these forces from their crystalline values $\delta f_{\parallel}$ and $\delta f_{\perp}$ can be thus be expressed as
\begin{eqnarray}
\nonumber
\delta f_{\parallel} &=& |f_{ij}| \cos(\theta_{ij}-\theta_{ij}^{0})-f_0,\\
\delta f_{\perp}   &=& |f_{ij}| \sin(\theta_{ij}-\theta_{ij}^{0}).
\end{eqnarray}
where $|f_{ij}|$ represents the magnitude of the force between particles $i$ and $j$, $\theta_{ij}$ is the bond angle between $i$ and $j$, and $\theta_{ij}^{(0)}$ is its value in the crystalline ordered state.
Using this decomposition we find that although both $|\delta f_x|$ and $|\delta f_y|$ have magnitudes of the same order, surprisingly the $\delta f_\perp$ caused by a single defect is much smaller in magnitude in comparison to $\delta f_\parallel$. To display this difference, in Fig.~\ref{Single_force_field} {\bf (a)} and {\bf (c)} we plot $\ln(|\delta f_\parallel|)$, and $\ln(|\delta f_\perp|)$ at each bond produced by a single defect from our simulations while in Fig.~\ref{Single_force_field} {\bf (b)} and Fig.~\ref{Single_force_field} {\bf (d)} we plot $\ln(|\delta f_\parallel|)$, and $\ln(|\delta f_\perp|)$ from the theoretically computed interparticle forces. This difference in the magnitude of the parallel and orthogonal components of forces produced by a single defect is the origin of constrained fluctuations in the orthogonal components of forces in minimally polydispersed disordered crystals~\cite{acharya2020athermal}. As superposition holds at linear order, the fluctuations in $\delta f_{\perp}$ produced by multiple defects, are also restricted. It should be noted that this effect only emerges by exactly solving {\it all} the equations of force balance simultaneously. 

\section{Large distance asymptotics}
\label{section_large_r}

\begin{figure}[t!]
\centering
\includegraphics[scale=0.4]{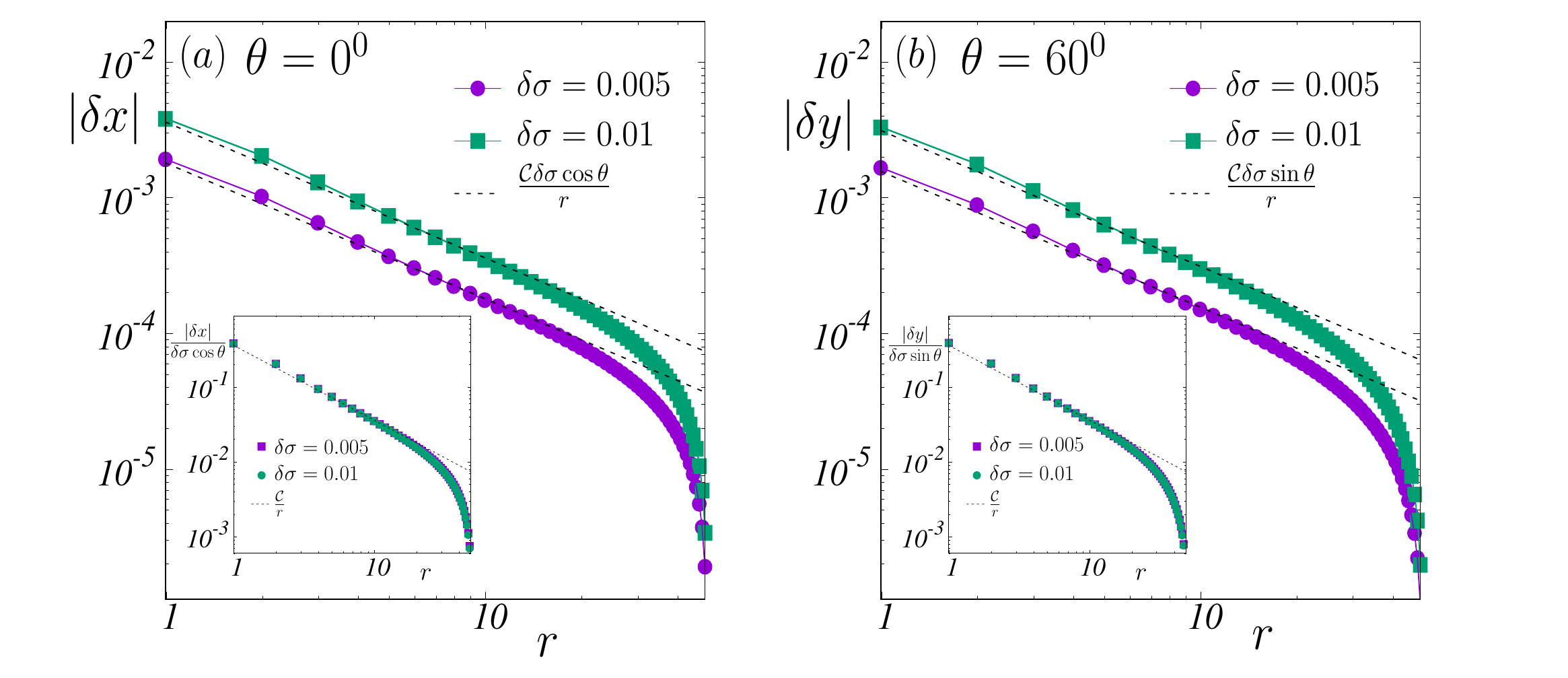}
\caption{
The displacement fields produced by a single defect for two different sizes of the defect, $\delta \sigma = 0.005$ and $\delta \sigma = 0.01$. The points represent the displacement fields obtained from simulations, and the solid lines represent our theoretical predictions at linear order.
{\bf (a)} The magnitude of the displacements fields $|\delta x|$, at distances measured along the $0^0$ angle with respect to the $x$-axis. ({\bf Inset}) These displacement fields scale linearly with $\delta \sigma$. {\bf (b)} The magnitude of displacement fields $|\delta y|$, at distances measured along the ${60}^0$ angle with respect to the $x$-axis. ({\bf Inset}) These displacement fields scale linearly with $\delta \sigma$.
The dashed lines represent our asymptotic predictions at large distances, given in Eq.~(\ref{eq_delx_large}). The prefactor is given by $\mathcal{C}= \frac{2 \sqrt{3} \delta \sigma  (1-\epsilon ) (1-2 \epsilon )}{\pi  (3-4 \epsilon )}$. Here we choose $\phi=0.92$ which sets $\epsilon = 7.14 \times 10^{-3}$ using Eq.~(\ref{eq_epsilon_definition}).
}
\label{collapse_xy} 
\end{figure} 

In this Section, we analyze the large distance behaviour of the solutions to the displacement fields. We focus on the linear order solutions and derive the continuum behaviour of the displacement field produced by a single defect in the system.
To obtain the large distance behaviour as $r \to \infty$, we analyze the solutions of the displacement fields in the small $k \to 0$ limit. 
It is convenient to work in radial coordinates in Fourier space $(\vec{k} = (k,\psi))$.
The Green's functions of the response given in Eq.~(\ref{eq_exact_green}) can be simplified in the small $k$ limit ($k\rightarrow 0$) to yield
\begin{eqnarray}
\nonumber
\Tilde{G}^{xx}(k,\psi,\epsilon) &=& -\frac{2 (\epsilon -1) (4 \epsilon  \cos (2 \psi )+8 \epsilon -3)}{k^2 \left(16 \epsilon
   ^2-16 \epsilon +3\right) (\cos (2 \psi )+2)^2},\\
\nonumber
\Tilde{G}^{xy}(k,\psi,\epsilon) &=& -\frac{2 \sqrt{3} (\epsilon -1) \sin (2 \psi )}{k^2 \left(16 \epsilon ^2-16 \epsilon
   +3\right) (\cos (2 \psi )+2)^2}, \\
   \nonumber
\Tilde{G}^{yy}(k,\psi,\epsilon) &=& -\frac{2 (\epsilon -1) (4 (\epsilon -1) \cos (2 \psi )+8 \epsilon -5)}{k^2 \left(16
   \epsilon ^2-16 \epsilon +3\right) (\cos (2 \psi )+2)^2},\\
\label{eq_smallk_green}
\end{eqnarray} 
with $\Tilde{G}^{y x} = \Tilde{G}^{x y}$. These Green's functions in Fourier space have the form $\Tilde{G}^{\mu\nu} = \frac{\Tilde{g}^{\mu\nu}(\psi,\epsilon)}{k^2}$. The prefactor $\Tilde{g}^{\mu\nu}(\psi,\epsilon)$ depends on the overcompression $\epsilon$ and hence on the packing fraction $\phi$. This asymptotic behaviour in Fourier space leads to a prediction of ${G}^{\mu\nu}(\vec{r}) \sim \log(|r|)$ for large distances $r$ for all the Green's functions in real space. These Green's functions in real space have been plotted in Fig.~\ref{Fig_green}. 

Next, the linear order source terms can be extracted from the functions $D^x$ and $D^y$ given in Eq.~(\ref{eq_D_functions}), which in the small $k$ limit can be simplified as
\begin{eqnarray}
\nonumber
D^{x}(k,\psi,\epsilon) &&= i 6(1-2\epsilon) k \cos \psi,\\
D^{y}(k,\psi,\epsilon) &&= i 2\sqrt{3}(1-2\epsilon) k \sin \psi.
\label{eq_smallk_dvec}
\end{eqnarray} 
Finally, using the expressions in Eqs.~(\ref{eq_smallk_green}) and (\ref{eq_smallk_dvec}) we arrive at the first order displacement fields in Fourier space in the small $k$ limit. We have
\begin{small}
\begin{eqnarray}
\nonumber
\delta \Tilde{x}^{(1)}(k,\psi,\epsilon) &=& i \delta \sigma \frac{12}{k} \frac{ (1-\epsilon ) (1-2 \epsilon ) }{ 3-4 \epsilon  } \frac{\cos (\psi )}{2 + \cos (2 \psi )},\\
\delta \Tilde{y}^{(1)}(k,\psi,\epsilon) &=& i \delta \sigma \frac{4 \sqrt{3}}{k} \frac{ (1-\epsilon ) (1-2 \epsilon ) }{ 3-4 \epsilon  } \frac{\sin (\psi )}{2 + \cos (2 \psi )}.
\label{eq_smallk_source}
\end{eqnarray} 
\end{small}
We next consider the large distance behaviour in the continuum limit, where $r/R_0\to \infty$, as well as $N\rightarrow\infty$. The displacement fields at large distances $r$ away from the defect can then be expressed as
\begin{eqnarray}
\nonumber
\delta x^{(1)}(r,\theta) = \frac{2}{4\pi^2}\int_{-\pi}^{\pi} d \psi \int_{0}^{\infty} \delta \Tilde{x}^{(1)}(k,\psi)\exp(-i\vec{k}.\vec{r}) k dk,\\
\nonumber
\delta y^{(1)}(r,\theta) = \frac{2}{4\pi^2}\int_{-\pi}^{\pi} d \psi \int_{0}^{\infty} \delta \Tilde{y}^{(1)}(k,\psi)\exp(-i\vec{k}.\vec{r}) k dk.\\
\label{eq_delx_cont}
\end{eqnarray}
The factors of $2$ that appear in the above expressions, represent the equal contributions to the integral from the singular behaviour of the Green's functions in Eq.~(\ref{eq_exact_green}) at the points $(0,0)$ and $(\pi,\pi)$. Finally, evaluating the above integrals using the expressions in Eq.~(\ref{eq_smallk_source}), we arrive at the following asymptotic forms of the linear order displacement fields at large distances
\begin{eqnarray}
\nonumber
 \delta x^{(1)}(r,\theta) &=& \mathcal{C} \delta\sigma \frac{\cos\theta}{r},\\
\delta y^{(1)}(r,\theta) &=& \mathcal{C} \delta\sigma \frac{\sin\theta}{r},
\label{eq_delx_large} 
\end{eqnarray} 
where the prefactor $\mathcal{C}$ depends on the overcompression $\epsilon$ and can be computed exactly. We have
\begin{equation}
\mathcal{C}= \frac{2 \sqrt{3}  (1-\epsilon ) (1-2 \epsilon )}{\pi  (3-4 \epsilon )}.
\end{equation}
\begin{figure}[t!]
\includegraphics[scale=0.4]{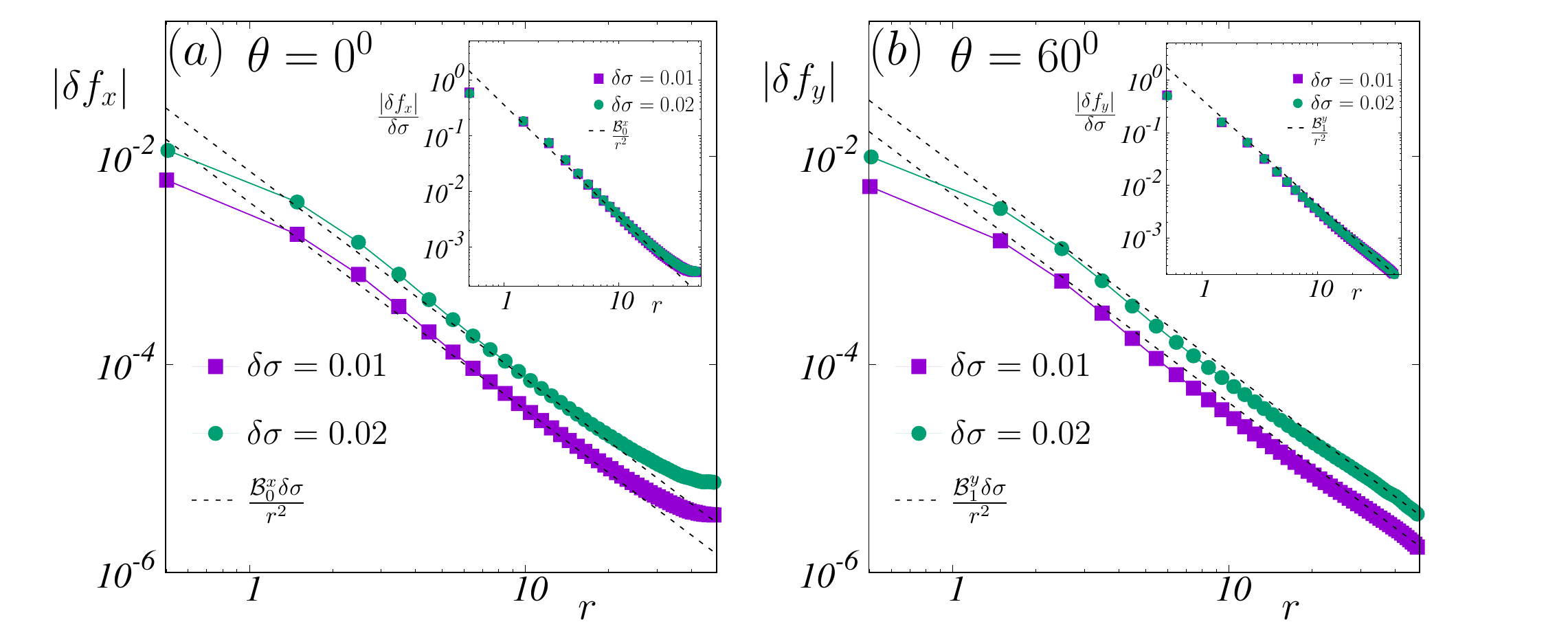}
\centering
\caption{The excess force at each bond produced by a single defect for two different sizes of the defect, $\delta \sigma = 0.005$ and $\delta \sigma = 0.01$. The points represent forces obtained from simulations, and the solid lines represent our theoretical predictions at linear order.
{\bf (a)} The magnitude of the excess force $|\delta f_x|$ at each bond, for distances measured along the $0^0$ angle with respect to the $x$-axis.  ({\bf Inset}) These excess force fields scale linearly with $\delta \sigma$. {\bf (b)} The magnitude of the excess force $|\delta f_y|$ at each bond, for distances measured along the ${60}^0$ angle with respect to the $x$-axis. ({\bf Inset}) These excess force fields scale linearly with $\delta \sigma$. The dashed lines represent our asymptotic predictions at large distances, given in Eq.~(\ref{eq_delf_large}).}
\label{rchange}
\end{figure}
We therefore find that in the continuum limit, the displacement fields at first order $\delta \vec{r}^{(1)}$ decay isotropically outwards from the defect as $\sim\frac{1}{r}$ at large distances. We note that the continuum limit solutions in Eq.~(\ref{eq_delx_large}) satisfy the Lam\'e equations in  Eq.~(\ref{eq_lame}). From the self similarity of the higher order solutions at large distances, we expect the continuum limits of the displacement fields at higher orders to also satisfy Eq.~(\ref{eq_lame}). This leads to corrections to the Lam\'e coefficients and consequently corrections to the elastic moduli that arise from the higher orders in our perturbation expansion, which in turn arise from the higher derivatives of the underlying interparticle potential. 
In Figs.~\ref{collapse_xy} {\bf (a)} and {\bf (b)} we plot the magnitude of the $x$-component of the displacement field along $\theta = 0^0$ and the $y$-component of the displacement field along $\theta = 60^0$, with increasing distance away from the defect. We plot these displacement fields for two different sizes of the defect $\delta \sigma = 0.005$ and $\delta\sigma = 0.01$. We also plot their scaling behaviour with $\delta \sigma$ along different angles $\theta$ in the insets of Figs.~\ref{collapse_xy} {\bf (a)} and {\bf (b)}, which are in agreement with our predictions in Eq.~(\ref{eq_delx_large}). We also display the convergence of these results to our continuum predictions in Eq.~(\ref{eq_delx_large}).

Next, the asymptotic behaviour of the excess forces can also be computed from the displacement field at large distance. Considering the forces between neighbouring particles $i$ and $j$ located at a large distance away from the defect, we have $r_i \simeq r_j \simeq r$ and $\theta_i \simeq \theta_j \simeq \theta$. Here  $r_i = |\vec{r}_i|$ is the distance of the particle from the defect, and $\theta_i$ is the angle of the distance vector $\vec{r}_i$ with respect to the $x$-axis.
Using Eq.~(\ref{eq_delx_large}), the relative displacements at linear order can be expressed as
 \begin{eqnarray}
 \nonumber
 \delta x^{(1)}_{ij}&=&\delta x_j - \delta x_i=\vec{\mathbb{r}}_j.\vec{\nabla}\delta x (\vec{r}),\\
 \delta y^{(1)}_{ij}&=&\delta y_j - \delta y_i=\vec{\mathbb{r}}_j.\vec{\nabla}\delta y (\vec{r}),
 \label{eq_cont_interparticle}
 \end{eqnarray}
where $\vec{\mathbb{r}}_j$ are the fundamental translation vectors of the lattice with $\vec{\mathbb{r}}_j = R_0(\cos \theta^{0}_{ij} \hat{x}+\sin \theta^{0}_{ij} \hat{y})$ and $\theta_{ij} = \theta_{j} - \theta_{i}$ and $\theta_{i}$. Using the expressions for the large distance behaviour of the displacement fields in Eq.~(\ref{eq_delx_large}), the relative displacements at linear order can be simplified to the following form
 \begin{eqnarray}
 \nonumber
 \delta x^{(1)}_{ij} (r,\theta)&=& \delta \sigma \frac{R_0 \mathcal{C }\cos\theta}{r^2}(\cos \theta^{0}_{ij}+\sin \theta^{0}_{ij}),\\
 \delta y^{(1)}_{ij} (r,\theta)&=& \delta \sigma \frac{R_0 \mathcal{C }\sin\theta}{r^2}(\cos \theta^{0}_{ij} +\sin \theta^{0}_{ij}).
 \label{eq_relative_displacement_asymptotics}
 \end{eqnarray}
We can next use these to compute the asymptotic behaviour of the change in interparticle forces produced by a single defect at linear order. Using the expressions in Eq.~(\ref{eq_relative_displacement_asymptotics}) in Eq.~(\ref{eq_forcebal_first_order}), we arrive at the excess forces at large distances away from the defect
\begin{small}
\begin{eqnarray}
\nonumber
\delta f^{x(1)}_{ij} (r,\theta)&=&  \mathcal{B}^{x}_j (\theta)  \frac{\delta \sigma}{r^2},\\
\delta f^{y(1)}_{ij} (r,\theta)&=&  \mathcal{B}^{y}_j (\theta)  \frac{\delta \sigma}{r^2},
\label{eq_delf_large}
\end{eqnarray}
\end{small}
where the constants are given by
\begin{small}
\begin{eqnarray}
 \nonumber
\mathcal{B}^{x}_j(\theta) &=&  R_0 \mathcal{C} (C_{ij}^{xx} \cos\theta + C_{ij}^{xy} \sin\theta)(\cos \theta^{0}_{ij}+\sin \theta^{0}_{ij}),\\
 \mathcal{B}^{y}_j(\theta) &=&  R_0 \mathcal{C} (C_{ij}^{yx} \cos\theta + C_{ij}^{yy} \sin\theta)(\cos \theta^{0}_{ij} +\sin \theta^{0}_{ij}).
 \label{eq_delf_coefficients}
\end{eqnarray}
\end{small}
In Fig.~\ref{rchange} {\bf (a)}, we plot the $|\delta f_x|$ along an angle $\theta=0^0$ with respect to $x$-axis for two different defect sizes $\delta \sigma = 0.005$ and $\delta\sigma = 0.01$. Our simulation results match with our predictions from the linear theory exactly, and display a $|\delta f^x|\sim1/r^2$ behaviour at large distances $r$ away from the defect. 
In Fig.~\ref{rchange} {\bf (b)}, we plot the $|\delta f_y|$ field at an angle $\theta=60^0$ with respect to the $x$-axis.
Once again these display a $|\delta f^y|\sim1/r^2$ behaviour at large distances $r$ from the defect. 
We also plot the scaling behaviour of these excess forces with $\delta \sigma$ along different angles $\theta$ in the insets of Figs.~\ref{rchange} {\bf (a)} and {\bf (b)}, which are in agreement with our theoretical predictions. We also display the convergence of these results to our continuum predictions in Eq.~(\ref{eq_delf_large}).
The finite perpendicular components of the interparticle forces produced by a single defect in the system can be attributed to the finite size of the lattice which appears through the vectors $\vec{\mathbb{r}}_{j}$ in Eq.~(\ref{eq_cont_interparticle}).
It is interesting to note that such an effect does not arise in continuum elasticity, as the solutions in Eq.~(\ref{eq_delx_large}) predict purely radial forces away from the defect in the continuum limit.

\subsection{Universal Behaviour}
\label{subsection_universal}

Many of our results derived for the system of polydispersed soft disks are universal, with the microscopic details of the model only changing the coefficients in the perturbation expansion developed in Section \ref{section_greens_function_expansion}. However, the underlying symmetries of the crystalline background persist in these coefficients, which could lead to differing behaviour at short length scales. 
While at small distances, the nature of the underlying lattice and the details of the force law are important, the large distance behaviour displays universal properties. 
As we have shown in Section \ref{section_large_r}, our displacement fields at linear order do not display the hallmarks of the underlying lattice at large lengthscales, displaying isotropic behaviour which also agrees with the predictions of continuum elasticity. However, as evidenced by the discussion of the higher-order coefficients $C_{ij}^{\alpha \beta \gamma}$ in Section \ref{section_greens_function_expansion}, these symmetries could play an important role in the corrections to universal behaviour.
The Green's function $G^{\mu\nu}(\vec{r})$ arising in our perturbation expansion can be derived for any central potential in a similar manner as in Section \ref{section_deformable_disks}. Additionally, since the solutions to the displacement fields at all orders involve the {\it same} Green's function, we can use the properties of the higher order sources to determine the universal scaling behaviour at higher orders. 
Since the displacement fields produced by a single defect decay with distance, these can be considered to be localized when the large lengthscale behaviour is considered. Therefore, the source terms at higher orders arising from a single defect are also localized.
This naturally leads to an explanation for the self similarity at all orders of the displacement fields arising from a single defect at large lengthscales.

The second order and higher order displacement fields produced by a single defect in the crystalline background also arise from localized sources, and are therefore expected to decay as $1/r$ at large distances with ``effective charges'' that control the magnitude at each order. Additionally, we expect these higher order displacement fields to be universal for all central potentials, with the displacement fields decaying as $1/r$ away from a defect, with the prefactors dependent on the underlying force law. Such behaviour can in turn be described by effective continuum equations at the coarse grained scale, which involve renormalized coefficients that include the higher terms in our perturbation theory. Although rotational invariance emerges at the large lengthscales away from a single defect in our theory, the higher order interactions between them encode the crystalline symmetries of the initial state, which we expect to persist in the large distance behaviour. This is in contrast to fully amorphous structures, where the lack of a reference state precludes the identification of such symmetries. We therefore expect rotational symmetries to emerge at large lengthscales in such systems as the microscopic disorder is increased. It would therefore be interesting to understand how our perturbation expansion breaks down as the disordered amorphous state is approached with larger amounts of disorder in the system.

\section{Higher order corrections arising from two defects}
\label{section_second_order_corrections}

Finally, we provide interesting avenues to test non-linear corrections to continuum elasticity in near-crystalline athermal systems. 
Having studied the displacement fields generated by a single defect in the crystalline background within our perturbation expansion, we turn to a non-trivial application of our theory, the case of two defects in a crystalline background. 
The second order corrections to the displacement fields produced by a single defect displayed in Fig.~\ref{Fig_displacement_field} can be obtained by subtracting the linear order solution from the numerically minimized displacement fields, as described in Section \ref{section_forces_single_defect}. Therefore such a quantity is not directly accessible through numerical simulations or experiments. 
We show below that such non-linear corrections can naturally be tested through numerically generated energy minimized configurations with {\it two} defects. The case of two defects therefore serves as an illustrative example of the non-trivial exact predictions of our theory.

The displacement fields generated by a single defect placed at positions $\vec{0}$ and $\vec{\Delta}$ respectively can be expressed in terms of our perturbation expansion as
\begin{eqnarray}
\nonumber
\delta \vec{r}_{\vec{0}} &=& \delta \vec{r}^{(1)}_{\vec{0}}+\delta \vec{r}^{(2)}_{\vec{0}}+\delta \vec{r}^{(3)}_{\vec{0}} + \hdots,\\
\delta \vec{r}_{\vec{\Delta}} &=& \delta \vec{r}^{(1)}_{\vec{\Delta}}+\delta \vec{r}^{(2)}_{\vec{\Delta}}+\delta \vec{r}^{(3)}_{\vec{\Delta}} + \hdots,
\end{eqnarray}
The LHS of the above equation represents the actual displacement fields that can be generated by an exact minimization of the energy. The RHS represents the series solution that can be extracted theoretically at each order.

\begin{figure}[t!]
\includegraphics[scale=0.4]{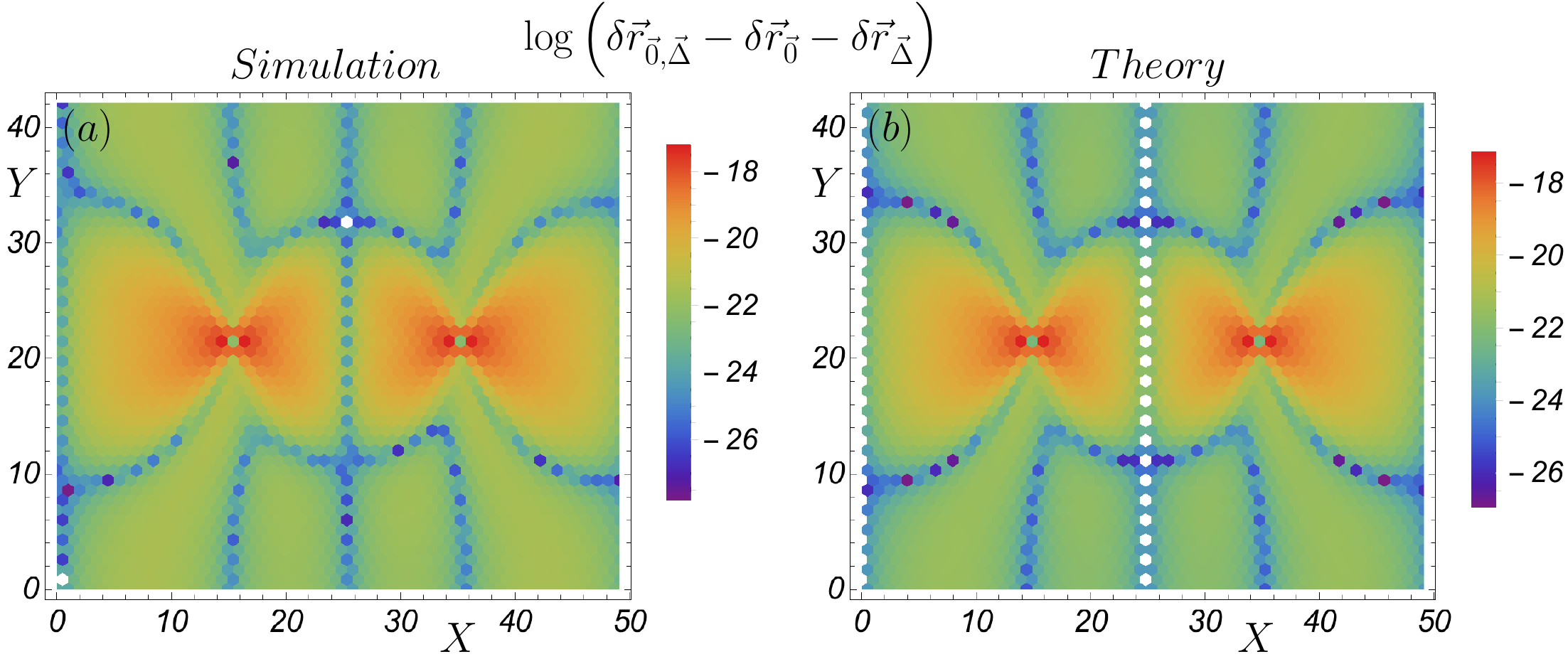}
\centering
\caption{Plot of the interaction in displacement fields due to two defects in the crystalline background $\delta \vec{r}_{\vec{0},\vec{\Delta}}- \delta \vec{r}_{\vec{0}}-\delta \vec{r}_{\vec{\Delta}}$. Here the field $\delta \vec{r}_{\vec{0},\vec{\Delta}}$ represents the displacements obtained by placing two defects at $(0,0)$ and $(\Delta,0)$ together, while $\delta \vec{r}_{\vec{0}}$ and $\delta \vec{r}_{\vec{\Delta}}$ are the individual displacement fields generated by placing single defects at $\vec{r}_{\vec{0}} = (0,0)$ and $\vec{r}_{\vec{\Delta}} = (\Delta,0)$ respectively. The magnitude of this interaction is small, which is due to the second order nature of this quantity. This naturally arises within our perturbation expansion, and matches with the numerically generated fields exactly. Here the system size is $N = 50$, and the defects are placed a distance $\Delta = 20$ apart along the $x$ direction. These configurations have been shifted using periodic boundary conditions to aid visualization. In our simulations in (a), we ensure that $|\sum_{j} \vec{f}_{ij}| \le 10^{-11}$ for each particle $i$, which is necessary in order to extract the sensitive corrections presented in (b).}
\label{fig_second_order_combination} 
\end{figure} 

Next, the displacement field arising from two defects placed together in the system at positions $(\vec{0})$ and $(\vec{\Delta})$ can be written in terms of our perturbation expansion as
\begin{eqnarray}
\delta \vec{r}_{\vec{0},\vec{\Delta}} &=& \delta \vec{r}^{(1)}_{\vec{0},\vec{\Delta}}+\delta \vec{r}^{(2)}_{\vec{0},\vec{\Delta}}+\delta \vec{r}^{(3)}_{\vec{0},\vec{\Delta}} + \hdots.
\end{eqnarray}
As is apparent from the Green's function formalism in Eq.~(\ref{eq_firstorder_disp_gen}), the displacement field arising from multiple defects at first order can be computed as a superposition of the displacement fields arising from each defect individually.
Therefore
\begin{equation}
\delta \vec{r}^{(1)}_{\vec{0},\vec{\Delta}} = \delta \vec{r}^{(1)}_{\vec{0}} + \delta \vec{r}^{(1)}_{\vec{\Delta}}.
\end{equation}
The second-order displacement fields arising from the two defects cannot be computed as the superposition of the displacement fields arising from each of them and therefore the second order fields for two defects cannot be described by the two individual fields alone. This non-linearily is encoded in the source terms at second order derived in Eq.~(\ref{eq_sec_source}) which involves a non-linear coupling of the linear order displacement fields. Therefore the second-order displacement fields arising from two defects together do not obey superposition. A non-trivial correction at second order can therefore be extracted from the combination
\begin{eqnarray}
\delta \vec{r}_{\vec{0},\vec{\Delta}}- \delta \vec{r}_{\vec{0}}-\delta \vec{r}_{\vec{\Delta}} &=& \delta \vec{r}^{(2)}_{\vec{0},\vec{\Delta}}- \delta \vec{r}^{(2)}_{\vec{0}}-\delta \vec{r}^{(2)}_{\vec{\Delta}}.
\label{eq_second_order_combination}
\end{eqnarray} 
Once again, the LHS and RHS of the above equation represent quantities that can be extracted numerically from energy minimized configurations and from our perturbation theory respectively. We note that the quantity in Eq.~(\ref{eq_second_order_combination}) arises purely at second order in our perturbation expansion. 
Such a prediction can be tested experimentally in systems where strains and forces can be measured, such as in photoelastic disks. We expect the effect become more prominent as one increases the size of the defect particles, with the intensity of the effect scaling as $\delta \sigma^2$. 

In Fig.~\ref{fig_second_order_combination} we plot the expression in Eq.~(\ref{eq_second_order_combination}) obtained from an exact numerical minimization of the system using the FIRE algorithm. We also plot the theoretical quantity in the RHS of Eq.~(\ref{eq_second_order_combination}) obtained from solving the displacement fields up to second order derived in Eqs.~(\ref{dis_k_green}) and (\ref{dis_k_green_second}).
This non-trivial interaction field obtained from simulations matches exactly with our theoretical predictions. We note that the interaction field in Eq.~(\ref{eq_second_order_combination}) predicted by our perturbation expansion cannot be captured within a continuum elasticity framework, as the corresponding Lam\'e equations in Eq.~(\ref{eq_lame}) are linear. Therefore such quantities can be used to experimentally test for deviations from linear elasticity in such athermal materials.

\section{Summary and conclusion}
\label{section_conclusion}
We have presented a new theoretical technique to exactly extract the displacement fields in disordered athermal systems near the crystalline state. Our hierarchical perturbation expansion about the crystalline state allows us to simultaneously solve the equations of force balance and obtain the displacement fields as a series expansion to arbitrary accuracy. We used this theory to study a physically relevant system of polydispersed soft disks, with disorder introduced in the radii of the particles. 
We illustrated our technique by exactly computing the displacement fields produced by a single defect introduced into the crystalline background, upto second order. These solutions match exactly with displacement fields extracted from our simulation of the energy minimized configuration with a single defect. Additionally, we illustrated the remarkable self-similar structure of these solutions at every order. Using our exact results, we derived a $|\delta r| \sim 1/r$ and $|\delta f| \sim 1/r^2$ decay for the displacement fields and excess forces at large distances $r$ away from the defect. Finally, we provided an experimentally testable prediction of our theory, the non-linear interaction between two defects, that can be extracted purely from the observed displacement fields produced by defects.

The perturbation expansion developed in this paper can easily be extended to involve systems with other types of microscopic disorder, such as spring networks with bond disorder, and also soft particle systems with external pinning forces~\cite{das2020long}. 
Our framework is quite general, and can also be applied to any precompressed crystalline system with particles interacting through central potentials. 
Our formalism can also be easily extended to different periodic backgrounds in two as well as higher dimensions.
In general solving the force balance equations involves the inversion of a large disordered matrix, which has up to now been difficult to formulate in terms of a coarse grained theory involving the properties of the underlying particles. In this regard our exact expansion techniques represent a useful analytic tool with which to develop coarse-grained theories for such materials. 


Since the equations governing continuum elasticity are linear, disordered athermal materials which display detectable non-linear effects, provide an interesting arena to search for higher order corrections to linear elasticity. 
Incorporating microscopic disorder along with force balance at the local level in disordered athermal materials continues to be a theoretical challenge. This is made possible in the paradigm of crystals, which allows an elegant representation of these microscopic degrees of freedom in Fourier space.  Since macroscopic constitutive equations are not well-defined for disordered amorphous solids, the formalism introduced in this paper that allows for exact predictions in the presence of microscopic disorder can be useful in understanding the emergence of constitutive equations at large lengthscales ~\cite{nampoothiri2020emergent}.



Since Green's function techniques find applications across various fields of physics, and are the building blocks of more complex emergent theories in continuum, it will be interesting to use our framework to develop non-linear elasticity theories describing athermal materials \cite{karmakar2010athermal}. It would also be interesting to develop re-summation techniques to simplify the higher order terms that appear in these expansions. Finally, our formulation can also be extended to incorporate disorder in the entire system, at linear order \cite{acharya2020athermal, das2020long}, as well as higher orders, to understand the emergence of amorphous properties in athermal systems with increasing disorder.

\subsection*{Acknowledgments}
We thank 
\mbox{Surajit} \mbox{Sengupta}, 
\mbox{Bulbul} \mbox{Chakraborty}, 
\mbox{Subhro} \mbox{Bhattacharjee}, 
\mbox{Mustansir} \mbox{Barma}, 
\mbox{Vishnu} V. \mbox{Krishnan} and 
\mbox{Jishnu} \mbox{Nampoothiri} 
for useful discussions. This project was funded by intramural funds at TIFR Hyderabad from the Department of Atomic Energy (DAE).


\bibliography{Disorder_Perturbation_Expansion_Bibliography}

\begin{thebibliography}{46}%
\makeatletter
\providecommand \@ifxundefined [1]{%
 \@ifx{#1\undefined}
}%
\providecommand \@ifnum [1]{%
 \ifnum #1\expandafter \@firstoftwo
 \else \expandafter \@secondoftwo
 \fi
}%
\providecommand \@ifx [1]{%
 \ifx #1\expandafter \@firstoftwo
 \else \expandafter \@secondoftwo
 \fi
}%
\providecommand \natexlab [1]{#1}%
\providecommand \enquote  [1]{``#1''}%
\providecommand \bibnamefont  [1]{#1}%
\providecommand \bibfnamefont [1]{#1}%
\providecommand \citenamefont [1]{#1}%
\providecommand \href@noop [0]{\@secondoftwo}%
\providecommand \href [0]{\begingroup \@sanitize@url \@href}%
\providecommand \@href[1]{\@@startlink{#1}\@@href}%
\providecommand \@@href[1]{\endgroup#1\@@endlink}%
\providecommand \@sanitize@url [0]{\catcode `\\12\catcode `\$12\catcode
  `\&12\catcode `\#12\catcode `\^12\catcode `\_12\catcode `\%12\relax}%
\providecommand \@@startlink[1]{}%
\providecommand \@@endlink[0]{}%
\providecommand \url  [0]{\begingroup\@sanitize@url \@url }%
\providecommand \@url [1]{\endgroup\@href {#1}{\urlprefix }}%
\providecommand \urlprefix  [0]{URL }%
\providecommand \Eprint [0]{\href }%
\providecommand \doibase [0]{https://doi.org/}%
\providecommand \selectlanguage [0]{\@gobble}%
\providecommand \bibinfo  [0]{\@secondoftwo}%
\providecommand \bibfield  [0]{\@secondoftwo}%
\providecommand \translation [1]{[#1]}%
\providecommand \BibitemOpen [0]{}%
\providecommand \bibitemStop [0]{}%
\providecommand \bibitemNoStop [0]{.\EOS\space}%
\providecommand \EOS [0]{\spacefactor3000\relax}%
\providecommand \BibitemShut  [1]{\csname bibitem#1\endcsname}%
\let\auto@bib@innerbib\@empty
\bibitem [{\citenamefont {Grigera}\ and\ \citenamefont
  {Israeloff}(1999)}]{grigera1999observation}%
  \BibitemOpen
  \bibfield  {author} {\bibinfo {author} {\bibfnamefont {T.~S.}\ \bibnamefont
  {Grigera}}\ and\ \bibinfo {author} {\bibfnamefont {N.}~\bibnamefont
  {Israeloff}},\ }\bibfield  {title} {\bibinfo {title} {Observation of
  fluctuation-dissipation-theorem violations in a structural glass},\
  }\href@noop {} {\bibfield  {journal} {\bibinfo  {journal} {Physical Review
  Letters}\ }\textbf {\bibinfo {volume} {83}},\ \bibinfo {pages} {5038}
  (\bibinfo {year} {1999})}\BibitemShut {NoStop}%
\bibitem [{\citenamefont {Landau}\ \emph {et~al.}(1986)\citenamefont {Landau},
  \citenamefont {Lifshitz} \emph {et~al.}}]{landau1987theoretical}%
  \BibitemOpen
  \bibfield  {author} {\bibinfo {author} {\bibfnamefont {L.~D.}\ \bibnamefont
  {Landau}}, \bibinfo {author} {\bibfnamefont {E.~M.}\ \bibnamefont
  {Lifshitz}}, \emph {et~al.},\ }\href@noop {} {\emph {\bibinfo {title} {Theory
  of elasticity}}},\ Vol.~\bibinfo {volume} {7}\ (\bibinfo  {publisher}
  {Pergamon Press, Oxford New York},\ \bibinfo {year} {1986})\BibitemShut
  {NoStop}%
\bibitem [{\citenamefont {Kubo}(1966)}]{kubo1966fluctuation}%
  \BibitemOpen
  \bibfield  {author} {\bibinfo {author} {\bibfnamefont {R.}~\bibnamefont
  {Kubo}},\ }\bibfield  {title} {\bibinfo {title} {The fluctuation-dissipation
  theorem},\ }\href@noop {} {\bibfield  {journal} {\bibinfo  {journal} {Reports
  on progress in physics}\ }\textbf {\bibinfo {volume} {29}},\ \bibinfo {pages}
  {255} (\bibinfo {year} {1966})}\BibitemShut {NoStop}%
\bibitem [{\citenamefont {Jaeger}\ \emph {et~al.}(1996)\citenamefont {Jaeger},
  \citenamefont {Nagel},\ and\ \citenamefont {Behringer}}]{jaeger1996granular}%
  \BibitemOpen
  \bibfield  {author} {\bibinfo {author} {\bibfnamefont {H.~M.}\ \bibnamefont
  {Jaeger}}, \bibinfo {author} {\bibfnamefont {S.~R.}\ \bibnamefont {Nagel}},\
  and\ \bibinfo {author} {\bibfnamefont {R.~P.}\ \bibnamefont {Behringer}},\
  }\bibfield  {title} {\bibinfo {title} {Granular solids, liquids, and gases},\
  }\href@noop {} {\bibfield  {journal} {\bibinfo  {journal} {Reviews of modern
  physics}\ }\textbf {\bibinfo {volume} {68}},\ \bibinfo {pages} {1259}
  (\bibinfo {year} {1996})}\BibitemShut {NoStop}%
\bibitem [{\citenamefont {van Hecke}(2009)}]{van2009jamming}%
  \BibitemOpen
  \bibfield  {author} {\bibinfo {author} {\bibfnamefont {M.}~\bibnamefont {van
  Hecke}},\ }\bibfield  {title} {\bibinfo {title} {Jamming of soft particles:
  geometry, mechanics, scaling and isostaticity},\ }\href@noop {} {\bibfield
  {journal} {\bibinfo  {journal} {Journal of Physics: Condensed Matter}\
  }\textbf {\bibinfo {volume} {22}},\ \bibinfo {pages} {033101} (\bibinfo
  {year} {2009})}\BibitemShut {NoStop}%
\bibitem [{\citenamefont {Henkes}\ \emph {et~al.}(2007)\citenamefont {Henkes},
  \citenamefont {O'Hern},\ and\ \citenamefont
  {Chakraborty}}]{henkes2007entropy}%
  \BibitemOpen
  \bibfield  {author} {\bibinfo {author} {\bibfnamefont {S.}~\bibnamefont
  {Henkes}}, \bibinfo {author} {\bibfnamefont {C.~S.}\ \bibnamefont {O'Hern}},\
  and\ \bibinfo {author} {\bibfnamefont {B.}~\bibnamefont {Chakraborty}},\
  }\bibfield  {title} {\bibinfo {title} {Entropy and temperature of a static
  granular assembly: An ab initio approach},\ }\href@noop {} {\bibfield
  {journal} {\bibinfo  {journal} {Physical review letters}\ }\textbf {\bibinfo
  {volume} {99}},\ \bibinfo {pages} {038002} (\bibinfo {year}
  {2007})}\BibitemShut {NoStop}%
\bibitem [{\citenamefont {O'Hern}\ \emph {et~al.}(2002)\citenamefont {O'Hern},
  \citenamefont {Langer}, \citenamefont {Liu},\ and\ \citenamefont
  {Nagel}}]{o2002random}%
  \BibitemOpen
  \bibfield  {author} {\bibinfo {author} {\bibfnamefont {C.~S.}\ \bibnamefont
  {O'Hern}}, \bibinfo {author} {\bibfnamefont {S.~A.}\ \bibnamefont {Langer}},
  \bibinfo {author} {\bibfnamefont {A.~J.}\ \bibnamefont {Liu}},\ and\ \bibinfo
  {author} {\bibfnamefont {S.~R.}\ \bibnamefont {Nagel}},\ }\bibfield  {title}
  {\bibinfo {title} {Random packings of frictionless particles},\ }\href@noop
  {} {\bibfield  {journal} {\bibinfo  {journal} {Physical Review Letters}\
  }\textbf {\bibinfo {volume} {88}},\ \bibinfo {pages} {075507} (\bibinfo
  {year} {2002})}\BibitemShut {NoStop}%
\bibitem [{\citenamefont {O'Hern}\ \emph {et~al.}(2003)\citenamefont {O'Hern},
  \citenamefont {Silbert}, \citenamefont {Liu},\ and\ \citenamefont
  {Nagel}}]{o2003jamming}%
  \BibitemOpen
  \bibfield  {author} {\bibinfo {author} {\bibfnamefont {C.~S.}\ \bibnamefont
  {O'Hern}}, \bibinfo {author} {\bibfnamefont {L.~E.}\ \bibnamefont {Silbert}},
  \bibinfo {author} {\bibfnamefont {A.~J.}\ \bibnamefont {Liu}},\ and\ \bibinfo
  {author} {\bibfnamefont {S.~R.}\ \bibnamefont {Nagel}},\ }\bibfield  {title}
  {\bibinfo {title} {Jamming at zero temperature and zero applied stress: The
  epitome of disorder},\ }\href@noop {} {\bibfield  {journal} {\bibinfo
  {journal} {Physical Review E}\ }\textbf {\bibinfo {volume} {68}},\ \bibinfo
  {pages} {011306} (\bibinfo {year} {2003})}\BibitemShut {NoStop}%
\bibitem [{\citenamefont {Wyart}(2005)}]{wyart2005rigidity}%
  \BibitemOpen
  \bibfield  {author} {\bibinfo {author} {\bibfnamefont {M.}~\bibnamefont
  {Wyart}},\ }\bibfield  {title} {\bibinfo {title} {On the rigidity of
  amorphous solids},\ }in\ \href@noop {} {\emph {\bibinfo {booktitle} {Annales
  de Physique}}},\ Vol.~\bibinfo {volume} {30}\ (\bibinfo {organization} {EDP
  Sciences},\ \bibinfo {year} {2005})\ pp.\ \bibinfo {pages}
  {1--96}\BibitemShut {NoStop}%
\bibitem [{\citenamefont {Goodrich}\ \emph {et~al.}(2012)\citenamefont
  {Goodrich}, \citenamefont {Liu},\ and\ \citenamefont
  {Nagel}}]{goodrich2012finite}%
  \BibitemOpen
  \bibfield  {author} {\bibinfo {author} {\bibfnamefont {C.~P.}\ \bibnamefont
  {Goodrich}}, \bibinfo {author} {\bibfnamefont {A.~J.}\ \bibnamefont {Liu}},\
  and\ \bibinfo {author} {\bibfnamefont {S.~R.}\ \bibnamefont {Nagel}},\
  }\bibfield  {title} {\bibinfo {title} {Finite-size scaling at the jamming
  transition},\ }\href@noop {} {\bibfield  {journal} {\bibinfo  {journal}
  {Physical review letters}\ }\textbf {\bibinfo {volume} {109}},\ \bibinfo
  {pages} {095704} (\bibinfo {year} {2012})}\BibitemShut {NoStop}%
\bibitem [{\citenamefont {Ramola}\ and\ \citenamefont
  {Chakraborty}(2017)}]{ramola2017scaling}%
  \BibitemOpen
  \bibfield  {author} {\bibinfo {author} {\bibfnamefont {K.}~\bibnamefont
  {Ramola}}\ and\ \bibinfo {author} {\bibfnamefont {B.}~\bibnamefont
  {Chakraborty}},\ }\bibfield  {title} {\bibinfo {title} {Scaling theory for
  the frictionless unjamming transition},\ }\href@noop {} {\bibfield  {journal}
  {\bibinfo  {journal} {Physical review letters}\ }\textbf {\bibinfo {volume}
  {118}},\ \bibinfo {pages} {138001} (\bibinfo {year} {2017})}\BibitemShut
  {NoStop}%
\bibitem [{\citenamefont {Cates}\ \emph {et~al.}(1998)\citenamefont {Cates},
  \citenamefont {Wittmer}, \citenamefont {Bouchaud},\ and\ \citenamefont
  {Claudin}}]{cates1998jamming}%
  \BibitemOpen
  \bibfield  {author} {\bibinfo {author} {\bibfnamefont {M.}~\bibnamefont
  {Cates}}, \bibinfo {author} {\bibfnamefont {J.}~\bibnamefont {Wittmer}},
  \bibinfo {author} {\bibfnamefont {J.-P.}\ \bibnamefont {Bouchaud}},\ and\
  \bibinfo {author} {\bibfnamefont {P.}~\bibnamefont {Claudin}},\ }\bibfield
  {title} {\bibinfo {title} {Jamming, force chains, and fragile matter},\
  }\href@noop {} {\bibfield  {journal} {\bibinfo  {journal} {Physical review
  letters}\ }\textbf {\bibinfo {volume} {81}},\ \bibinfo {pages} {1841}
  (\bibinfo {year} {1998})}\BibitemShut {NoStop}%
\bibitem [{\citenamefont {Torquato}\ and\ \citenamefont
  {Stillinger}(2010)}]{torquato2010jammed}%
  \BibitemOpen
  \bibfield  {author} {\bibinfo {author} {\bibfnamefont {S.}~\bibnamefont
  {Torquato}}\ and\ \bibinfo {author} {\bibfnamefont {F.~H.}\ \bibnamefont
  {Stillinger}},\ }\bibfield  {title} {\bibinfo {title} {Jammed hard-particle
  packings: From kepler to bernal and beyond},\ }\href@noop {} {\bibfield
  {journal} {\bibinfo  {journal} {Reviews of modern physics}\ }\textbf
  {\bibinfo {volume} {82}},\ \bibinfo {pages} {2633} (\bibinfo {year}
  {2010})}\BibitemShut {NoStop}%
\bibitem [{\citenamefont {Berthier}\ and\ \citenamefont
  {Biroli}(2011)}]{berthier2011theoretical}%
  \BibitemOpen
  \bibfield  {author} {\bibinfo {author} {\bibfnamefont {L.}~\bibnamefont
  {Berthier}}\ and\ \bibinfo {author} {\bibfnamefont {G.}~\bibnamefont
  {Biroli}},\ }\bibfield  {title} {\bibinfo {title} {Theoretical perspective on
  the glass transition and amorphous materials},\ }\href@noop {} {\bibfield
  {journal} {\bibinfo  {journal} {Reviews of Modern Physics}\ }\textbf
  {\bibinfo {volume} {83}},\ \bibinfo {pages} {587} (\bibinfo {year}
  {2011})}\BibitemShut {NoStop}%
\bibitem [{\citenamefont {Kapteijns}\ \emph {et~al.}(2019)\citenamefont
  {Kapteijns}, \citenamefont {Ji}, \citenamefont {Brito}, \citenamefont
  {Wyart},\ and\ \citenamefont {Lerner}}]{kapteijns2019fast}%
  \BibitemOpen
  \bibfield  {author} {\bibinfo {author} {\bibfnamefont {G.}~\bibnamefont
  {Kapteijns}}, \bibinfo {author} {\bibfnamefont {W.}~\bibnamefont {Ji}},
  \bibinfo {author} {\bibfnamefont {C.}~\bibnamefont {Brito}}, \bibinfo
  {author} {\bibfnamefont {M.}~\bibnamefont {Wyart}},\ and\ \bibinfo {author}
  {\bibfnamefont {E.}~\bibnamefont {Lerner}},\ }\bibfield  {title} {\bibinfo
  {title} {Fast generation of ultrastable computer glasses by minimization of
  an augmented potential energy},\ }\href@noop {} {\bibfield  {journal}
  {\bibinfo  {journal} {Physical Review E}\ }\textbf {\bibinfo {volume} {99}},\
  \bibinfo {pages} {012106} (\bibinfo {year} {2019})}\BibitemShut {NoStop}%
\bibitem [{\citenamefont {Bi}\ \emph {et~al.}(2011)\citenamefont {Bi},
  \citenamefont {Zhang}, \citenamefont {Chakraborty},\ and\ \citenamefont
  {Behringer}}]{bi2011jamming}%
  \BibitemOpen
  \bibfield  {author} {\bibinfo {author} {\bibfnamefont {D.}~\bibnamefont
  {Bi}}, \bibinfo {author} {\bibfnamefont {J.}~\bibnamefont {Zhang}}, \bibinfo
  {author} {\bibfnamefont {B.}~\bibnamefont {Chakraborty}},\ and\ \bibinfo
  {author} {\bibfnamefont {R.~P.}\ \bibnamefont {Behringer}},\ }\bibfield
  {title} {\bibinfo {title} {Jamming by shear},\ }\href@noop {} {\bibfield
  {journal} {\bibinfo  {journal} {Nature}\ }\textbf {\bibinfo {volume} {480}},\
  \bibinfo {pages} {355} (\bibinfo {year} {2011})}\BibitemShut {NoStop}%
\bibitem [{\citenamefont {Goodrich}\ \emph {et~al.}(2014)\citenamefont
  {Goodrich}, \citenamefont {Liu},\ and\ \citenamefont {Nagel}}]{Goodrich2014}%
  \BibitemOpen
  \bibfield  {author} {\bibinfo {author} {\bibfnamefont {C.~P.}\ \bibnamefont
  {Goodrich}}, \bibinfo {author} {\bibfnamefont {A.~J.}\ \bibnamefont {Liu}},\
  and\ \bibinfo {author} {\bibfnamefont {S.~R.}\ \bibnamefont {Nagel}},\
  }\bibfield  {title} {\bibinfo {title} {Solids between the mechanical extremes
  of order and disorder},\ }\href@noop {} {\bibfield  {journal} {\bibinfo
  {journal} {Nature Physics}\ }\textbf {\bibinfo {volume} {10}},\ \bibinfo
  {pages} {578} (\bibinfo {year} {2014})}\BibitemShut {NoStop}%
\bibitem [{\citenamefont {Tong}\ \emph {et~al.}(2015)\citenamefont {Tong},
  \citenamefont {Tan},\ and\ \citenamefont {Xu}}]{tong2015crystals}%
  \BibitemOpen
  \bibfield  {author} {\bibinfo {author} {\bibfnamefont {H.}~\bibnamefont
  {Tong}}, \bibinfo {author} {\bibfnamefont {P.}~\bibnamefont {Tan}},\ and\
  \bibinfo {author} {\bibfnamefont {N.}~\bibnamefont {Xu}},\ }\bibfield
  {title} {\bibinfo {title} {From crystals to disordered crystals: A hidden
  order-disorder transition},\ }\href@noop {} {\bibfield  {journal} {\bibinfo
  {journal} {Scientific reports}\ }\textbf {\bibinfo {volume} {5}},\ \bibinfo
  {pages} {15378} (\bibinfo {year} {2015})}\BibitemShut {NoStop}%
\bibitem [{\citenamefont {Acharya}\ \emph {et~al.}(2020)\citenamefont
  {Acharya}, \citenamefont {Sengupta}, \citenamefont {Chakraborty},\ and\
  \citenamefont {Ramola}}]{acharya2020athermal}%
  \BibitemOpen
  \bibfield  {author} {\bibinfo {author} {\bibfnamefont {P.}~\bibnamefont
  {Acharya}}, \bibinfo {author} {\bibfnamefont {S.}~\bibnamefont {Sengupta}},
  \bibinfo {author} {\bibfnamefont {B.}~\bibnamefont {Chakraborty}},\ and\
  \bibinfo {author} {\bibfnamefont {K.}~\bibnamefont {Ramola}},\ }\bibfield
  {title} {\bibinfo {title} {Athermal fluctuations in disordered crystals},\
  }\href@noop {} {\bibfield  {journal} {\bibinfo  {journal} {Physical Review
  Letters}\ }\textbf {\bibinfo {volume} {124}},\ \bibinfo {pages} {168004}
  (\bibinfo {year} {2020})}\BibitemShut {NoStop}%
\bibitem [{\citenamefont {Eshelby}(1956)}]{eshelby1956continuum}%
  \BibitemOpen
  \bibfield  {author} {\bibinfo {author} {\bibfnamefont {J.}~\bibnamefont
  {Eshelby}},\ }\bibfield  {title} {\bibinfo {title} {The continuum theory of
  lattice defects},\ }\href@noop {} {\bibfield  {journal} {\bibinfo  {journal}
  {Solid state physics}\ }\textbf {\bibinfo {volume} {3}},\ \bibinfo {pages}
  {79} (\bibinfo {year} {1956})}\BibitemShut {NoStop}%
\bibitem [{\citenamefont {Batra}(1987)}]{batra1987force}%
  \BibitemOpen
  \bibfield  {author} {\bibinfo {author} {\bibfnamefont {R.}~\bibnamefont
  {Batra}},\ }\bibfield  {title} {\bibinfo {title} {The force on a lattice
  defect in an elastic body},\ }\href@noop {} {\bibfield  {journal} {\bibinfo
  {journal} {Journal of elasticity}\ }\textbf {\bibinfo {volume} {17}},\
  \bibinfo {pages} {3} (\bibinfo {year} {1987})}\BibitemShut {NoStop}%
\bibitem [{\citenamefont {Eshelby}(1957)}]{eshelby1957determination}%
  \BibitemOpen
  \bibfield  {author} {\bibinfo {author} {\bibfnamefont {J.~D.}\ \bibnamefont
  {Eshelby}},\ }\bibfield  {title} {\bibinfo {title} {The determination of the
  elastic field of an ellipsoidal inclusion, and related problems},\
  }\href@noop {} {\bibfield  {journal} {\bibinfo  {journal} {Proceedings of the
  royal society of London. Series A. Mathematical and physical sciences}\
  }\textbf {\bibinfo {volume} {241}},\ \bibinfo {pages} {376} (\bibinfo {year}
  {1957})}\BibitemShut {NoStop}%
\bibitem [{\citenamefont {Duan}\ \emph {et~al.}(2005)\citenamefont {Duan},
  \citenamefont {Wang}, \citenamefont {Huang},\ and\ \citenamefont
  {Karihaloo}}]{duan2005eshelby}%
  \BibitemOpen
  \bibfield  {author} {\bibinfo {author} {\bibfnamefont {H.}~\bibnamefont
  {Duan}}, \bibinfo {author} {\bibfnamefont {J.}~\bibnamefont {Wang}}, \bibinfo
  {author} {\bibfnamefont {Z.}~\bibnamefont {Huang}},\ and\ \bibinfo {author}
  {\bibfnamefont {B.~L.}\ \bibnamefont {Karihaloo}},\ }\bibfield  {title}
  {\bibinfo {title} {Eshelby formalism for nano-inhomogeneities},\ }\href@noop
  {} {\bibfield  {journal} {\bibinfo  {journal} {Proceedings of the Royal
  Society A: Mathematical, Physical and Engineering Sciences}\ }\textbf
  {\bibinfo {volume} {461}},\ \bibinfo {pages} {3335} (\bibinfo {year}
  {2005})}\BibitemShut {NoStop}%
\bibitem [{\citenamefont {Nampoothiri}\ \emph {et~al.}(2020)\citenamefont
  {Nampoothiri}, \citenamefont {Wang}, \citenamefont {Ramola}, \citenamefont
  {Zhang}, \citenamefont {Bhattacharjee},\ and\ \citenamefont
  {Chakraborty}}]{nampoothiri2020emergent}%
  \BibitemOpen
  \bibfield  {author} {\bibinfo {author} {\bibfnamefont {J.~N.}\ \bibnamefont
  {Nampoothiri}}, \bibinfo {author} {\bibfnamefont {Y.}~\bibnamefont {Wang}},
  \bibinfo {author} {\bibfnamefont {K.}~\bibnamefont {Ramola}}, \bibinfo
  {author} {\bibfnamefont {J.}~\bibnamefont {Zhang}}, \bibinfo {author}
  {\bibfnamefont {S.}~\bibnamefont {Bhattacharjee}},\ and\ \bibinfo {author}
  {\bibfnamefont {B.}~\bibnamefont {Chakraborty}},\ }\bibfield  {title}
  {\bibinfo {title} {Emergent elasticity in amorphous solids},\ }\href@noop {}
  {\bibfield  {journal} {\bibinfo  {journal} {Physical review letters}\
  }\textbf {\bibinfo {volume} {125}},\ \bibinfo {pages} {118002} (\bibinfo
  {year} {2020})}\BibitemShut {NoStop}%
\bibitem [{\citenamefont {Makse}\ and\ \citenamefont
  {Kurchan}(2002)}]{makse2002testing}%
  \BibitemOpen
  \bibfield  {author} {\bibinfo {author} {\bibfnamefont {H.~A.}\ \bibnamefont
  {Makse}}\ and\ \bibinfo {author} {\bibfnamefont {J.}~\bibnamefont
  {Kurchan}},\ }\bibfield  {title} {\bibinfo {title} {Testing the thermodynamic
  approach to granular matter with a numerical model of a decisive
  experiment},\ }\href@noop {} {\bibfield  {journal} {\bibinfo  {journal}
  {Nature}\ }\textbf {\bibinfo {volume} {415}},\ \bibinfo {pages} {614}
  (\bibinfo {year} {2002})}\BibitemShut {NoStop}%
\bibitem [{\citenamefont {Tighe}\ \emph {et~al.}(2010)\citenamefont {Tighe},
  \citenamefont {Snoeijer}, \citenamefont {Vlugt},\ and\ \citenamefont {van
  Hecke}}]{tighe2010force}%
  \BibitemOpen
  \bibfield  {author} {\bibinfo {author} {\bibfnamefont {B.~P.}\ \bibnamefont
  {Tighe}}, \bibinfo {author} {\bibfnamefont {J.~H.}\ \bibnamefont {Snoeijer}},
  \bibinfo {author} {\bibfnamefont {T.~J.}\ \bibnamefont {Vlugt}},\ and\
  \bibinfo {author} {\bibfnamefont {M.}~\bibnamefont {van Hecke}},\ }\bibfield
  {title} {\bibinfo {title} {The force network ensemble for granular
  packings},\ }\href@noop {} {\bibfield  {journal} {\bibinfo  {journal} {Soft
  Matter}\ }\textbf {\bibinfo {volume} {6}},\ \bibinfo {pages} {2908} (\bibinfo
  {year} {2010})}\BibitemShut {NoStop}%
\bibitem [{\citenamefont {Edwards}\ and\ \citenamefont
  {Oakeshott}(1989)}]{edwards1989theory}%
  \BibitemOpen
  \bibfield  {author} {\bibinfo {author} {\bibfnamefont {S.~F.}\ \bibnamefont
  {Edwards}}\ and\ \bibinfo {author} {\bibfnamefont {R.}~\bibnamefont
  {Oakeshott}},\ }\bibfield  {title} {\bibinfo {title} {Theory of powders},\
  }\href@noop {} {\bibfield  {journal} {\bibinfo  {journal} {Physica A:
  Statistical Mechanics and its Applications}\ }\textbf {\bibinfo {volume}
  {157}},\ \bibinfo {pages} {1080} (\bibinfo {year} {1989})}\BibitemShut
  {NoStop}%
\bibitem [{\citenamefont {Blumenfeld}\ and\ \citenamefont
  {Edwards}(2009)}]{blumenfeld2009granular}%
  \BibitemOpen
  \bibfield  {author} {\bibinfo {author} {\bibfnamefont {R.}~\bibnamefont
  {Blumenfeld}}\ and\ \bibinfo {author} {\bibfnamefont {S.~F.}\ \bibnamefont
  {Edwards}},\ }\bibfield  {title} {\bibinfo {title} {On granular stress
  statistics: Compactivity, angoricity, and some open issues},\ }\href@noop {}
  {\bibfield  {journal} {\bibinfo  {journal} {The Journal of Physical Chemistry
  B}\ }\textbf {\bibinfo {volume} {113}},\ \bibinfo {pages} {3981} (\bibinfo
  {year} {2009})}\BibitemShut {NoStop}%
\bibitem [{\citenamefont {Liu}\ \emph {et~al.}(1995)\citenamefont {Liu},
  \citenamefont {Nagel}, \citenamefont {Schecter}, \citenamefont {Coppersmith},
  \citenamefont {Majumdar}, \citenamefont {Narayan},\ and\ \citenamefont
  {Witten}}]{liu1995force}%
  \BibitemOpen
  \bibfield  {author} {\bibinfo {author} {\bibfnamefont {C.-h.}\ \bibnamefont
  {Liu}}, \bibinfo {author} {\bibfnamefont {S.~R.}\ \bibnamefont {Nagel}},
  \bibinfo {author} {\bibfnamefont {D.}~\bibnamefont {Schecter}}, \bibinfo
  {author} {\bibfnamefont {S.}~\bibnamefont {Coppersmith}}, \bibinfo {author}
  {\bibfnamefont {S.}~\bibnamefont {Majumdar}}, \bibinfo {author}
  {\bibfnamefont {O.}~\bibnamefont {Narayan}},\ and\ \bibinfo {author}
  {\bibfnamefont {T.}~\bibnamefont {Witten}},\ }\bibfield  {title} {\bibinfo
  {title} {Force fluctuations in bead packs},\ }\href@noop {} {\bibfield
  {journal} {\bibinfo  {journal} {Science}\ }\textbf {\bibinfo {volume}
  {269}},\ \bibinfo {pages} {513} (\bibinfo {year} {1995})}\BibitemShut
  {NoStop}%
\bibitem [{\citenamefont {Coppersmith}\ \emph {et~al.}(1996)\citenamefont
  {Coppersmith}, \citenamefont {Liu}, \citenamefont {Majumdar}, \citenamefont
  {Narayan},\ and\ \citenamefont {Witten}}]{coppersmith1996model}%
  \BibitemOpen
  \bibfield  {author} {\bibinfo {author} {\bibfnamefont {S.}~\bibnamefont
  {Coppersmith}}, \bibinfo {author} {\bibfnamefont {C.-h.}\ \bibnamefont
  {Liu}}, \bibinfo {author} {\bibfnamefont {S.}~\bibnamefont {Majumdar}},
  \bibinfo {author} {\bibfnamefont {O.}~\bibnamefont {Narayan}},\ and\ \bibinfo
  {author} {\bibfnamefont {T.}~\bibnamefont {Witten}},\ }\bibfield  {title}
  {\bibinfo {title} {Model for force fluctuations in bead packs},\ }\href@noop
  {} {\bibfield  {journal} {\bibinfo  {journal} {Physical Review E}\ }\textbf
  {\bibinfo {volume} {53}},\ \bibinfo {pages} {4673} (\bibinfo {year}
  {1996})}\BibitemShut {NoStop}%
\bibitem [{\citenamefont {Das}\ \emph {et~al.}(2020)\citenamefont {Das},
  \citenamefont {Acharya},\ and\ \citenamefont {Ramola}}]{das2020long}%
  \BibitemOpen
  \bibfield  {author} {\bibinfo {author} {\bibfnamefont {D.}~\bibnamefont
  {Das}}, \bibinfo {author} {\bibfnamefont {P.}~\bibnamefont {Acharya}},\ and\
  \bibinfo {author} {\bibfnamefont {K.}~\bibnamefont {Ramola}},\ }\bibfield
  {title} {\bibinfo {title} {Long-range correlations in actively pinned
  athermal networks},\ }\href@noop {} {\bibfield  {journal} {\bibinfo
  {journal} {arXiv preprint arXiv:2012.08541}\ } (\bibinfo {year}
  {2020})}\BibitemShut {NoStop}%
\bibitem [{\citenamefont {Theil}(2006)}]{theil2006proof}%
  \BibitemOpen
  \bibfield  {author} {\bibinfo {author} {\bibfnamefont {F.}~\bibnamefont
  {Theil}},\ }\bibfield  {title} {\bibinfo {title} {A proof of crystallization
  in two dimensions},\ }\href@noop {} {\bibfield  {journal} {\bibinfo
  {journal} {Communications in Mathematical Physics}\ }\textbf {\bibinfo
  {volume} {262}},\ \bibinfo {pages} {209} (\bibinfo {year}
  {2006})}\BibitemShut {NoStop}%
\bibitem [{\citenamefont {Wang}\ \emph {et~al.}(2018)\citenamefont {Wang},
  \citenamefont {Ding}, \citenamefont {Yan}, \citenamefont {Asta},
  \citenamefont {Ritchie},\ and\ \citenamefont {Li}}]{wang2018spatial}%
  \BibitemOpen
  \bibfield  {author} {\bibinfo {author} {\bibfnamefont {N.}~\bibnamefont
  {Wang}}, \bibinfo {author} {\bibfnamefont {J.}~\bibnamefont {Ding}}, \bibinfo
  {author} {\bibfnamefont {F.}~\bibnamefont {Yan}}, \bibinfo {author}
  {\bibfnamefont {M.}~\bibnamefont {Asta}}, \bibinfo {author} {\bibfnamefont
  {R.~O.}\ \bibnamefont {Ritchie}},\ and\ \bibinfo {author} {\bibfnamefont
  {L.}~\bibnamefont {Li}},\ }\bibfield  {title} {\bibinfo {title} {Spatial
  correlation of elastic heterogeneity tunes the deformation behavior of
  metallic glasses},\ }\href@noop {} {\bibfield  {journal} {\bibinfo  {journal}
  {npj Computational Materials}\ }\textbf {\bibinfo {volume} {4}},\ \bibinfo
  {pages} {1} (\bibinfo {year} {2018})}\BibitemShut {NoStop}%
\bibitem [{\citenamefont {Ma}\ and\ \citenamefont
  {Ding}(2016)}]{ma2016tailoring}%
  \BibitemOpen
  \bibfield  {author} {\bibinfo {author} {\bibfnamefont {E.}~\bibnamefont
  {Ma}}\ and\ \bibinfo {author} {\bibfnamefont {J.}~\bibnamefont {Ding}},\
  }\bibfield  {title} {\bibinfo {title} {Tailoring structural inhomogeneities
  in metallic glasses to enable tensile ductility at room temperature},\
  }\href@noop {} {\bibfield  {journal} {\bibinfo  {journal} {Materials Today}\
  }\textbf {\bibinfo {volume} {19}},\ \bibinfo {pages} {568} (\bibinfo {year}
  {2016})}\BibitemShut {NoStop}%
\bibitem [{\citenamefont {Zaccone}\ \emph {et~al.}(2007)\citenamefont
  {Zaccone}, \citenamefont {Lattuada}, \citenamefont {Wu},\ and\ \citenamefont
  {Morbidelli}}]{zaccone2007theoretical}%
  \BibitemOpen
  \bibfield  {author} {\bibinfo {author} {\bibfnamefont {A.}~\bibnamefont
  {Zaccone}}, \bibinfo {author} {\bibfnamefont {M.}~\bibnamefont {Lattuada}},
  \bibinfo {author} {\bibfnamefont {H.}~\bibnamefont {Wu}},\ and\ \bibinfo
  {author} {\bibfnamefont {M.}~\bibnamefont {Morbidelli}},\ }\bibfield  {title}
  {\bibinfo {title} {Theoretical elastic moduli for disordered packings of
  interconnected spheres},\ }\href@noop {} {\bibfield  {journal} {\bibinfo
  {journal} {The Journal of chemical physics}\ }\textbf {\bibinfo {volume}
  {127}},\ \bibinfo {pages} {174512} (\bibinfo {year} {2007})}\BibitemShut
  {NoStop}%
\bibitem [{\citenamefont {Agnolin}\ and\ \citenamefont
  {Kruyt}(2008)}]{agnolin2008elastic}%
  \BibitemOpen
  \bibfield  {author} {\bibinfo {author} {\bibfnamefont {I.}~\bibnamefont
  {Agnolin}}\ and\ \bibinfo {author} {\bibfnamefont {N.~P.}\ \bibnamefont
  {Kruyt}},\ }\bibfield  {title} {\bibinfo {title} {On the elastic moduli of
  two-dimensional assemblies of disks: Relevance and modeling of fluctuations
  in particle displacements and rotations},\ }\href@noop {} {\bibfield
  {journal} {\bibinfo  {journal} {Computers \& Mathematics with Applications}\
  }\textbf {\bibinfo {volume} {55}},\ \bibinfo {pages} {245} (\bibinfo {year}
  {2008})}\BibitemShut {NoStop}%
\bibitem [{\citenamefont {Watanabe}\ and\ \citenamefont
  {Tanaka}(2008)}]{watanabe2008direct}%
  \BibitemOpen
  \bibfield  {author} {\bibinfo {author} {\bibfnamefont {K.}~\bibnamefont
  {Watanabe}}\ and\ \bibinfo {author} {\bibfnamefont {H.}~\bibnamefont
  {Tanaka}},\ }\bibfield  {title} {\bibinfo {title} {Direct observation of
  medium-range crystalline order in granular liquids near the glass
  transition},\ }\href@noop {} {\bibfield  {journal} {\bibinfo  {journal}
  {Physical review letters}\ }\textbf {\bibinfo {volume} {100}},\ \bibinfo
  {pages} {158002} (\bibinfo {year} {2008})}\BibitemShut {NoStop}%
\bibitem [{\citenamefont {Hu}\ \emph {et~al.}(2017)\citenamefont {Hu},
  \citenamefont {Tanaka},\ and\ \citenamefont {Wang}}]{hu2017impact}%
  \BibitemOpen
  \bibfield  {author} {\bibinfo {author} {\bibfnamefont {Y.-C.}\ \bibnamefont
  {Hu}}, \bibinfo {author} {\bibfnamefont {H.}~\bibnamefont {Tanaka}},\ and\
  \bibinfo {author} {\bibfnamefont {W.-H.}\ \bibnamefont {Wang}},\ }\bibfield
  {title} {\bibinfo {title} {Impact of spatial dimension on structural ordering
  in metallic glass},\ }\href@noop {} {\bibfield  {journal} {\bibinfo
  {journal} {Physical Review E}\ }\textbf {\bibinfo {volume} {96}},\ \bibinfo
  {pages} {022613} (\bibinfo {year} {2017})}\BibitemShut {NoStop}%
\bibitem [{\citenamefont {Tah}\ \emph {et~al.}(2018)\citenamefont {Tah},
  \citenamefont {Sengupta}, \citenamefont {Sastry}, \citenamefont {Dasgupta},\
  and\ \citenamefont {Karmakar}}]{tah2018glass}%
  \BibitemOpen
  \bibfield  {author} {\bibinfo {author} {\bibfnamefont {I.}~\bibnamefont
  {Tah}}, \bibinfo {author} {\bibfnamefont {S.}~\bibnamefont {Sengupta}},
  \bibinfo {author} {\bibfnamefont {S.}~\bibnamefont {Sastry}}, \bibinfo
  {author} {\bibfnamefont {C.}~\bibnamefont {Dasgupta}},\ and\ \bibinfo
  {author} {\bibfnamefont {S.}~\bibnamefont {Karmakar}},\ }\bibfield  {title}
  {\bibinfo {title} {Glass transition in supercooled liquids with medium-range
  crystalline order},\ }\href@noop {} {\bibfield  {journal} {\bibinfo
  {journal} {Physical review letters}\ }\textbf {\bibinfo {volume} {121}},\
  \bibinfo {pages} {085703} (\bibinfo {year} {2018})}\BibitemShut {NoStop}%
\bibitem [{\citenamefont {Mecking}(1979)}]{mecking1979deformation}%
  \BibitemOpen
  \bibfield  {author} {\bibinfo {author} {\bibfnamefont {H.}~\bibnamefont
  {Mecking}},\ }\bibfield  {title} {\bibinfo {title} {Deformation of
  polycrystals},\ }in\ \href@noop {} {\emph {\bibinfo {booktitle} {Strength of
  metals and alloys}}}\ (\bibinfo  {publisher} {Elsevier},\ \bibinfo {year}
  {1979})\ pp.\ \bibinfo {pages} {1573--1594}\BibitemShut {NoStop}%
\bibitem [{\citenamefont {Sastry}\ \emph {et~al.}(1971)\citenamefont {Sastry},
  \citenamefont {Prasad},\ and\ \citenamefont {Vasu}}]{sastry1971low}%
  \BibitemOpen
  \bibfield  {author} {\bibinfo {author} {\bibfnamefont {D.}~\bibnamefont
  {Sastry}}, \bibinfo {author} {\bibfnamefont {Y.}~\bibnamefont {Prasad}},\
  and\ \bibinfo {author} {\bibfnamefont {K.}~\bibnamefont {Vasu}},\ }\bibfield
  {title} {\bibinfo {title} {Low-temperature deformation behaviour of
  polycrystalline copper},\ }\href@noop {} {\bibfield  {journal} {\bibinfo
  {journal} {Journal of Materials Science}\ }\textbf {\bibinfo {volume} {6}},\
  \bibinfo {pages} {1433} (\bibinfo {year} {1971})}\BibitemShut {NoStop}%
\bibitem [{\citenamefont {Yiannourakou}\ \emph {et~al.}(2010)\citenamefont
  {Yiannourakou}, \citenamefont {Economou},\ and\ \citenamefont
  {Bitsanis}}]{yiannourakou2010structural}%
  \BibitemOpen
  \bibfield  {author} {\bibinfo {author} {\bibfnamefont {M.}~\bibnamefont
  {Yiannourakou}}, \bibinfo {author} {\bibfnamefont {I.~G.}\ \bibnamefont
  {Economou}},\ and\ \bibinfo {author} {\bibfnamefont {I.~A.}\ \bibnamefont
  {Bitsanis}},\ }\bibfield  {title} {\bibinfo {title} {Structural and dynamical
  analysis of monodisperse and polydisperse colloidal systems},\ }\href@noop {}
  {\bibfield  {journal} {\bibinfo  {journal} {The Journal of chemical physics}\
  }\textbf {\bibinfo {volume} {133}},\ \bibinfo {pages} {224901} (\bibinfo
  {year} {2010})}\BibitemShut {NoStop}%
\bibitem [{\citenamefont {Wigner}(1934)}]{wigner1934interaction}%
  \BibitemOpen
  \bibfield  {author} {\bibinfo {author} {\bibfnamefont {E.}~\bibnamefont
  {Wigner}},\ }\bibfield  {title} {\bibinfo {title} {On the interaction of
  electrons in metals},\ }\href@noop {} {\bibfield  {journal} {\bibinfo
  {journal} {Physical Review}\ }\textbf {\bibinfo {volume} {46}},\ \bibinfo
  {pages} {1002} (\bibinfo {year} {1934})}\BibitemShut {NoStop}%
\bibitem [{\citenamefont {Horiguchi}(1972)}]{horiguchi1972lattice}%
  \BibitemOpen
  \bibfield  {author} {\bibinfo {author} {\bibfnamefont {T.}~\bibnamefont
  {Horiguchi}},\ }\bibfield  {title} {\bibinfo {title} {Lattice green's
  functions for the triangular and honeycomb lattices},\ }\href@noop {}
  {\bibfield  {journal} {\bibinfo  {journal} {Journal of Mathematical Physics}\
  }\textbf {\bibinfo {volume} {13}},\ \bibinfo {pages} {1411} (\bibinfo {year}
  {1972})}\BibitemShut {NoStop}%
\bibitem [{\citenamefont {Durian}(1995)}]{durian1995foam}%
  \BibitemOpen
  \bibfield  {author} {\bibinfo {author} {\bibfnamefont {D.~J.}\ \bibnamefont
  {Durian}},\ }\bibfield  {title} {\bibinfo {title} {Foam mechanics at the
  bubble scale},\ }\href@noop {} {\bibfield  {journal} {\bibinfo  {journal}
  {Physical review letters}\ }\textbf {\bibinfo {volume} {75}},\ \bibinfo
  {pages} {4780} (\bibinfo {year} {1995})}\BibitemShut {NoStop}%
\bibitem [{\citenamefont {Karmakar}\ \emph {et~al.}(2010)\citenamefont
  {Karmakar}, \citenamefont {Lerner},\ and\ \citenamefont
  {Procaccia}}]{karmakar2010athermal}%
  \BibitemOpen
  \bibfield  {author} {\bibinfo {author} {\bibfnamefont {S.}~\bibnamefont
  {Karmakar}}, \bibinfo {author} {\bibfnamefont {E.}~\bibnamefont {Lerner}},\
  and\ \bibinfo {author} {\bibfnamefont {I.}~\bibnamefont {Procaccia}},\
  }\bibfield  {title} {\bibinfo {title} {Athermal nonlinear elastic constants
  of amorphous solids},\ }\href@noop {} {\bibfield  {journal} {\bibinfo
  {journal} {Physical Review E}\ }\textbf {\bibinfo {volume} {82}},\ \bibinfo
  {pages} {026105} (\bibinfo {year} {2010})}\BibitemShut {NoStop}%
\end{thebibliography}%


\clearpage
\begin{appendix}

\begin{widetext}

\section*{Expansion coefficients for polydispersed disks}

The linear order coefficients in the perturbation expansion for polydispersed disks can be expressed as (setting $K=1$)
\begin{small}
\begin{eqnarray}
\nonumber
C^{xx}_{ij}(R_0) &=& -\frac{R_0 - \sigma_0 + \sigma_0 \cos(\frac{2 \pi j}{3})}{4 R_0 \sigma_0^2},\\
\nonumber
C^{xy}_{ij}(R_0) &=& -\frac{\sin(\frac{2\pi j}{3})}{4 R_0 \sigma_0}  ,\\
\nonumber
C^{x \sigma}_{ij}(R_0) &=& \frac{\Big(R_0 - \sigma_0\Big)\cos(\frac{\pi j}{3})}{4\sigma_0^3},\\
\nonumber
C^{yx}_{ij}(R_0) &=& \frac{\sin(\frac{2\pi j}{3})}{4 R_0 \sigma_0},\\
\nonumber
C^{yy}_{ij}(R_0) &=&  \frac{\sigma_0 - R_0 + \sigma_0 \cos(\frac{2 \pi j}{3})}{4 R_0 \sigma_0^2},\\
\nonumber
C^{y \sigma}_{ij}(\phi) &=& \frac{\Big(R_0 - \sigma_0\Big) \sin(\frac{2\pi j}{3})}{4 \sigma_0^3}.
\end{eqnarray}
\end{small}

Since the second order coefficients in the perturbation expansion require an antisymmetrization over indices, it is convenient to define the index $\overline{j} = \textrm{mod}(j + 3,6)$, representing the neighbour of particle $i$ in the opposite direction to $j$ (see Fig.~\ref{fig_lattice_notation}). In this notation, we have
\begin{small}
\begin{eqnarray}
\nonumber
C^{xxx}_{ij}(R_0) &=& 
\frac{R_0 \left(\cos(\frac{2\overline{j}\pi}{3}) - \cos(\frac{2j\pi}{3})\right) + \frac{3}{4} \left(\cos(\frac{\overline{j}\pi}{3}) - \cos(\overline{j}\pi) -\cos(\frac{j\pi}{3}) + \cos(j\pi)\right)}{8R_0^2 \sigma_0} \\
\nonumber
C^{xxy}_{ij}(R_0) &=& \frac{\sin(\frac{\pi\overline{j}}{3})- 3\sin(\pi \overline{j})- \sin(\frac{\pi j}{3}) + 3\sin(\pi j)} {16 R_0^2\sigma_0},\\
\nonumber
C^{xyy}_{ij}(R_0) &=& \frac{\cos(\frac{\pi \overline{j}}{3}) + 3\cos(\pi \overline{j}) + 4 R_0 \sin(\frac{2\pi\overline{j}}{3}) - \cos(\frac{\pi j}{3}) - 3\cos(\pi j) - 4 R_0 \sin(\frac{2\pi j}{3})}{32 R_0^2 \sigma_0},\\
\nonumber
C^{xx\sigma}_{ij}(R_0) &=& \frac{4R_0 - 2\sigma_0 + \sigma_0\left(\cos(\frac{2\pi\overline{j}}{3}) + \cos(\frac{2\pi j}{3})\right)}{16 R_0 \sigma_0^3},\\
\nonumber
C^{xy\sigma}_{ij}(\phi) &=& \frac{\sin(\frac{2 \pi \overline{j}}{3}) -\sin(\frac{2 \pi j}{3})}{16 R_0 \sigma_0^2},\\
\nonumber
C^{x\sigma\sigma}_{ij}(\phi) &=& \frac{(3 - 2\sigma_0)\left(\cos(\frac{\pi \overline{j}}{3}) - \cos(\frac{\pi j}{3})\right)}{32\sigma_0^4},\\
\nonumber
C^{yxx}_{ij}(R_0) &=& \frac{\sin(\frac{\pi \overline{j}}{3}) - 3\sin(\pi \overline{j}) + 4 R_0 \sin(\frac{2\pi\overline{j}}{3}) - \sin(\frac{\pi j}{3}) + 3\sin(\pi j) - 4 R_0 \sin(\frac{2\pi j}{3})}{32 R_0^2 \sigma_0},\\
\nonumber
C^{yxy}_{ij}(R_0) &=& -\frac{\cos(\frac{\pi \overline{j}}{3})+ 3\cos(\pi \overline{j}) - \cos(\frac{\pi j}{3})- 3\cos(\pi j)}{16 R_0^2 \sigma_0},\\
\nonumber
C^{yyy}_{ij}(R_0) &=& -\frac{4 R_0 \left(\cos(\frac{2\pi j}{3})- \cos(\frac{2\pi\overline{j}}{3})\right) - \left(\sin(\frac{\pi j}{3}) + \sin(\pi j) - \sin(\frac{\pi \overline{j}}{3}) - \sin(\pi \overline{j})\right)}{32 R_0^2 \sigma_0},\\
\nonumber
C^{yx\sigma}_{ij}(R_0) &=& \frac{\sin(\frac{2 \pi j}{3})+ \sin(\frac{2 \pi \overline{j}}{3})}{16 R_0\sigma_0^2},\\
\nonumber
C^{yy\sigma}_{ij}(R_0) &=& \frac{-4R_0 + 2\sigma_0 + \sigma_0\left(\cos(\frac{2\pi\overline{j}}{3}) + \cos(\frac{2\pi j}{3})\right)}{16 R_0 \sigma_0^3},\\
\nonumber
C^{y\sigma\sigma}_{ij}(R_0) &=& \frac{(3 - 2\sigma_0)\left(\sin(\frac{\pi \overline{j}}{3}) - \sin(\frac{\pi j}{3})\right)}{32\sigma_0^4},\\
\end{eqnarray}
\end{small}

\clearpage

\end{widetext}

\end{appendix}

\end{document}